\documentclass[12pt]{article}
\usepackage[a4paper,margin=1in]{geometry}
\usepackage{amsmath,amssymb}
\usepackage{graphicx}
\usepackage{booktabs}
\usepackage{siunitx}
\usepackage{caption}
\usepackage{subcaption}
\usepackage{hyperref}
\usepackage{cleveref}
\usepackage[numbers]{natbib}
\usepackage{float}
\usepackage{xcolor}

\title{Explainable Patterns in Cryptocurrency Microstructure}
\author{
  Bartosz Bieganowski\thanks{bbieganowsk@student.uw.edu.pl, https://orcid.org/0009-0003-8671-8266} \\
  \small
  University of Warsaw, Faculty of Economic Sciences \\
  \small
  Department of Quantitative Finance and Machine Learning\\
  \small
  Quantiative Finance Research Group
  \and
  Robert Ślepaczuk, PhD, DSc (Habil.), prof. UW\thanks{rslepaczuk@wne.uw.edu.pl, https://orcid.org/0000-0001-5227-2014} \\
  \small
  University of Warsaw, Faculty of Economic Sciences \\
  \small
  Department of Quantitative Finance and Machine Learning\\
  \small
  Quantiative Finance Research Group
}
\date{\today}

\graphicspath{{charts/BTC/}{charts/ENJ/}{charts/ETC/}{charts/LTC/}{charts/ROSE/}{figures/}}

\newcommand{\rts}{r_{t\to t+3\,s}}
\newcommand{\midp}{\mathrm{mid}}

\begin{document}
\maketitle
\begin{abstract}
We document stable cross-asset patterns in cryptocurrency limit-order-book microstructure: the same engineered order book and trade features exhibit remarkably similar predictive importance and SHAP dependence shapes across assets spanning an order of magnitude in market capitalization (BTC, LTC, ETC, ENJ, ROSE). The data covers Binance Futures perpetual contract order books and trades on 1-second frequency starting from January 1st, 2022 up to October 12th, 2025. Using a unified CatBoost modeling pipeline with a direction-aware GMADL objective and time-series cross validation, we show that feature rankings and partial effects are stable across assets despite heterogeneous liquidity and volatility. We connect these SHAP structures to microstructure theory (order flow imbalance, spread, and adverse selection) and validate tradability via a conservative top-of-book taker backtest as well as fixed depth maker backtest. Our primary novelty is a robustness analysis of a major flash crash, where the divergent performance of our taker and maker strategies empirically validates classic microstructure theories of adverse selection and highlights the systemic risks of algorithmic trading. Our results suggest a portable microstructure representation of short-horizon returns and motivate universal feature libraries for crypto markets.

\end{abstract}

\section{Introduction}
Financial market microstructure studies how trading, information, and liquidity provision jointly determine short-horizon price dynamics. Across asset classes, a robust set of features (order flow imbalance, bid–ask spreads, depth, and trade arrival patterns) has been shown to explain a substantial fraction of return variation at very short horizons. Cryptocurrencies offer a unique space for testing whether these features are universal: assets vary widely in capitalization and liquidity, yet they are transacted through similar continuous double-auction mechanisms with transparent limit order books.

This paper advances the hypothesis that short-horizon return predictability in crypto admits a universal representation. Concretely, we posit that a compact set of features engineered from the top of the order book and contemporaneous trade flow exhibits similar predictive importance and functional dependence shapes across assets spanning an order of magnitude in market capitalization. We evaluate this claim with flexible non-linear models and model-agnostic explanation tools designed for tabular data.

Our empirical strategy is deliberately simple and portable. Given tick-level orderbook and trades data for five cryptocurrencies, we engineer a unified feature library capturing spreads and relative prices, order flow imbalances, and deviations of buy/sell volume-weighted average prices from the mid. We train gradient-boosted decision trees with a direction-aware objective tailored to financial returns, alongside a standard squared-error baseline. Out-of-sample explanations are computed using SHAP values to obtain both global importance rankings and local partial-dependence-like curves.

Three findings emerge. First, the same families of features dominate the SHAP summaries across large-cap and mid/long-tail cryptoassets. Second, SHAP dependence shapes are strikingly consistent across assets: order flow imbalance has a largely monotone effect with concavity at extremes; spreads are associated with diminished predictability; and VWAP-to-mid deviations display asymmetric effects coherent with short-lived pressure and microstructure reversion. Third, when coupled with a conservative taker execution that marks inventory on the unfavorable side of the book, the forecasts translate into tradable signals under reasonable thresholds, indicating economic significance rather than mere statistical regularities.

Our work offers two primary novel contributions to the literature. First, we move beyond simple feature importance rankings by conducting an in-depth, cross-asset analysis of SHAP dependence shapes, connecting these empirical patterns to established microstructure theory. This provides a more granular understanding of how information is impounded into prices. Second, we present a unique robustness analysis by stress-testing our models during a major market flash crash. The divergent performance of our taker (liquidity-demanding) and maker (liquidity-providing) strategies during this event serves as a powerful empirical validation of classic adverse selection theories and offers a cautionary tale on the systemic risks inherent in modern, algorithmically-driven markets.

Conceptually, these results support a scale-invariant view of microstructure: once normalized into relative prices and flows, the mechanisms by which liquidity-taking pressure and information asymmetry affect short-horizon returns appear stable across capitalization tiers. Practically, the findings motivate the use of universal feature libraries and reduce the need for asset-specific feature engineering when deploying short-horizon models across the crypto universe.

The remainder of the paper proceeds as follows. We situate our study within the microstructure and explainable machine learning literatures, describe the data and engineered features, detail the modeling and evaluation methodology, analyze SHAP-based explanations and their theoretical interpretations, assess tradability via a conservative taker backtest, and discuss robustness checks and implications before concluding.

\section{Literature Review}
A large literature links trading frictions, information, and prices. Foundational inventory- and information-based models establish the theoretical link between trading activity and price formation. The seminal model of \citet{Kyle1985} shows how an informed insider strategically places trades over time, and how market makers infer information from aggregate order flow to set prices. Complementing this, \citet{GlostenMilgrom1985} model the bid-ask spread as a direct consequence of adverse selection, where liquidity providers widen their quotes to compensate for the risk of trading against privately informed agents. These theories, synthesized empirically by \citet{Hasbrouck2007}, provide the basis for our features related to order flow and spreads. A more recent, data-driven perspective from econophysics, particularly the work of \citet{Bouchaud2009} and \citet{Cont2011}, documents universal statistical patterns in price impact, such as the "square-root" law, suggesting that endogenous feedback loops in the trading process itself are as important as fundamentals. This motivates our use of flexible, non-linear models capable of capturing these complex dynamics.

Limit order book research has evolved to document rich predictive structures. Early work by \citet{Gould2013} provides a comprehensive review of LOB mechanics, while more recent approaches leverage deep learning. \citet{Sirignano2019} demonstrate universal features of price formation using large-scale deep learning models on limit order book data, finding that spatial and temporal dependencies are consistent across asset classes. Similarly, \citet{Zhang2019} propose DeepLOB, applying convolutional neural networks to extract features from raw order book states. Recent studies published in \textit{Expert Systems with Applications} further validate the efficacy of machine learning in this domain. \citet{Briola2024} introduce deep learning architectures for forecasting limit order books that capture complex temporal dependencies. \citet{Prata2023} conduct a benchmark study of LOB-based deep learning models, confirming their robustness in trend prediction. On the feature engineering side, \citet{Stoikov2018} introduces the micro-price, a high-frequency estimator of future prices that adjusts the mid-price based on bid-ask imbalances, a concept directly related to the imbalance features we evaluate. Our work bridges these streams by applying interpretable tree-based models to engineered features, aiming to recover the universal patterns found by deep learning but with the transparency required for economic validation.

Our methodology is grounded in explainable machine learning (XAI) to transparently connect these complex data patterns to economic theory. We employ CatBoost \citep{CatBoost2018}, a state-of-the-art gradient boosting algorithm optimized for the kind of tabular data prevalent in microstructure, known for its strong performance without requiring extensive data preprocessing. To interpret these potentially opaque models, we rely on SHAP (SHapley Additive exPlanations), a game-theoretic framework that explains model predictions. The foundational work of \citet{LundbergLee2017} unified various methods into a single, theoretically sound approach, and we specifically use the efficient TreeSHAP algorithm tailored for tree-based models like CatBoost \citep{Lundberg2020}. By using SHAP, we can decompose individual predictions into feature contributions, allowing us to validate whether the model has learned relationships consistent with microstructure theory.

Crypto microstructure has converged toward traditional double-auction dynamics while retaining distinctive features (heterogeneous participants, varying tick sizes, fragmented liquidity). Evidence indicates that order flow imbalance, spreads, and liquidity shocks are predictive of near-term returns and execution costs \citep{MakarovSchoar2020}. \citet{Chen2021} demonstrate the utility of machine learning in predicting Bitcoin exchange rates using a broad set of economic and technical determinants. What remains less explored is the extent to which predictive features and their functional effects are universal across the crypto capitalization spectrum.

Recent work has addressed the challenge of cross-validation in financial time series, where standard k-fold approaches can fail due to temporal dependencies and selection bias. Walk-forward validation procedures have been proposed for evaluating machine learning models on financial assets, with evidence supporting the relative effectiveness of LSTM and GRU architectures with respect to more basic recurrent architectures \citep{BaranochnikovSlepaczuk2022}. \citet{Ong2023} further extend this by applying deep multi-task learning to construct time-series momentum portfolios, highlighting the importance of robust validation schemes. To further mitigate overlapping information between training and test sets, combinatorial purged cross-validation methods have been introduced, explicitly accounting for the non-IID nature of financial returns \citep{DePrado2018}. \citet{Shabani2023} also contribute to this area by proposing augmented bilinear networks for robust financial time series prediction.

The impact of loss function choice on high-frequency prediction and trading performance has also been explored using deep learning architectures tailored to time-series data. Informer-based models trained on high-frequency Bitcoin trades have been benchmarked across a variety of loss functions, including RMSE, quantile loss, and the recently introduced Generalized Mean Absolute Directional Loss (GMADL) \citep{MADL}. Notably, the GMADL objective demonstrated robust improvement in both predictive accuracy and trading outcomes, especially on finer time intervals, underscoring the importance of direction-aware loss functions for short-horizon forecasting in liquid markets \citep{StefaniukSlepaczuk2025}.

We contribute by testing a universality hypothesis with SHAP-based diagnostics: that both the ranking of feature importances and the shapes of their marginal effects are stable across assets with disparate market caps. This complements prior work on impact functions and liquidity provision by emphasizing cross-asset invariance in the mapping from microstructure features to short-horizon returns.

\section{Data}
The dataset consists of multiple cryptocurrencies spanning a range of liquidity conditions, aiming to capture the diversity of market microstructure effects. At the start of the dataset period (January 1, 2022), these assets occupied market capitalization ranks 1, 20, 40, 60, and 100, respectively\footnote{\url{https://coinmarketcap.com/historical/20220102/}}, providing a broad cross-section from the most liquid to smaller-cap coins. We construct synchronized time series for each asset, pairing sub-second top-of-book quotes with trade records to obtain a detailed view of the order book dynamics and trading activity. The data is sourced from Binance Futures perpetual contract order books and trades on 1-second frequency starting from January 1st, 2022 up to October 12th, 2025. 

Our prediction target is the logarithmic return of the mid price over a short horizon, specifically:

\begin{equation}
    \rts = \log(\midp_{t+3\,s}/\midp_t) \text{ where } \midp_t = \frac{ask_0 + bid_0}{2}
\end{equation}

Log returns are preferred in financial modeling because they are symmetric in terms of impact on equity returns for positive and negative moves, additive over time, and naturally account for compounding, which simplifies statistical analysis. In addition, this choice reduces sensitivity to large price changes in absolute terms and facilitates comparison across assets with very different price levels. Using a short prediction horizon helps to isolate microstructural effects while minimizing issues related to overlapping information or delayed market reactions.

The features selected for modeling can be grouped into several intuitive families, each motivated by specific a priori hypotheses about market behavior:
\begin{itemize}
    \item \textbf{Top-of-book metrics} (such as the mid price, spread, and level 1 volumes) are designed to reflect immediate liquidity and trading costs. For example, a wider spread may indicate deteriorated liquidity or greater uncertainty, potentially foreshadowing price movements. High or unbalanced top-of-book volumes might signal supply and demand imbalances at the best available prices.
    \item \textbf{Order flow and trade imbalance features} (e.g., net traded volume or measures like signed order flow) aim to capture the ongoing pressure exerted by aggressive traders. Persistent buying or selling is often associated with short-term market impact, as liquidity providers adjust quotes in response to observed order flow.
    \item \textbf{Deviations of VWAP from the mid price} (for both buy and sell trades) are included to quantify how recent trading activity has occurred at prices away from the midpoint, which can be indicative of latent pressure or the presence of informed trading. Large deviations might point to short-term adverse selection or one-sided interest in the order book.
\end{itemize}
Deep order book levels and redundant aggregates are omitted to avoid introducing noise and to maintain feature sets that are robust and interpretable across different markets.

In terms of preprocessing, feature construction is aligned precisely in time using forward filling for short-lived gaps to mirror the real-time availability of order book data. Any samples missing essential information are excluded from the analysis. Importantly, we deliberately retain each feature in its original scale and units. This is because our chosen models (ensemble tree methods) are insensitive to monotonic transformations and can internally learn appropriate splits or thresholds regardless of feature scaling. As a result, explicit rescaling or normalization is unnecessary, and the model remains free to discover the most relevant functional relationships.

Relative measures such as spreads-to-mid or VWAP-to-mid deviations are preferred as input features whenever possible. These ratios are inherently scale-invariant, supporting direct comparison of feature effects across assets, even when absolute price levels or tick sizes differ substantially. This comparability is especially important when interpreting global feature importances or dependence plots produced by model explanations such as SHAP.

\begin{figure}[h]
  \centering
  \begin{subfigure}{0.32\textwidth}
    \includegraphics[width=\linewidth]{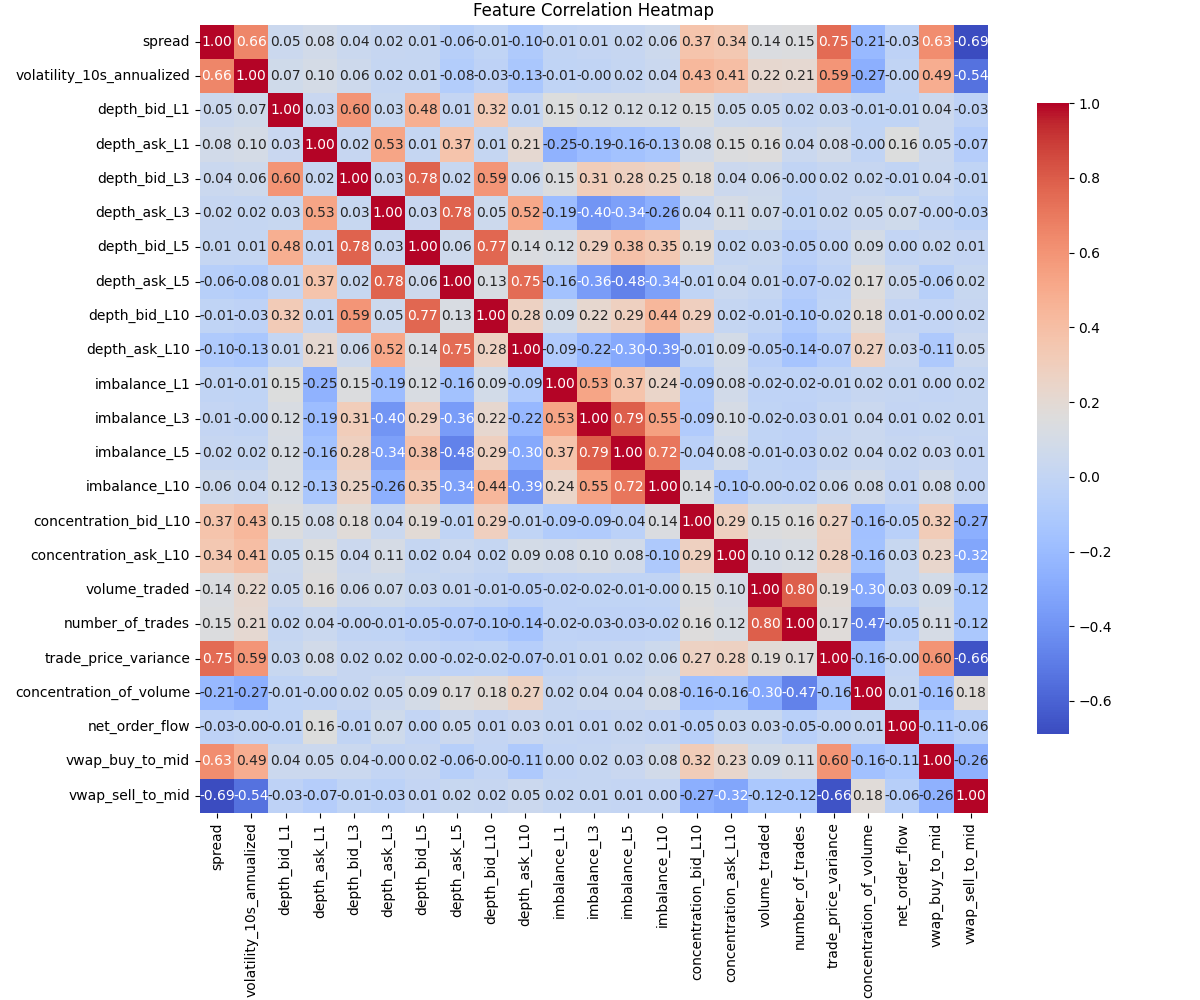}
    \caption{ENJ}
  \end{subfigure}
  \begin{subfigure}{0.32\textwidth}
    \includegraphics[width=\linewidth]{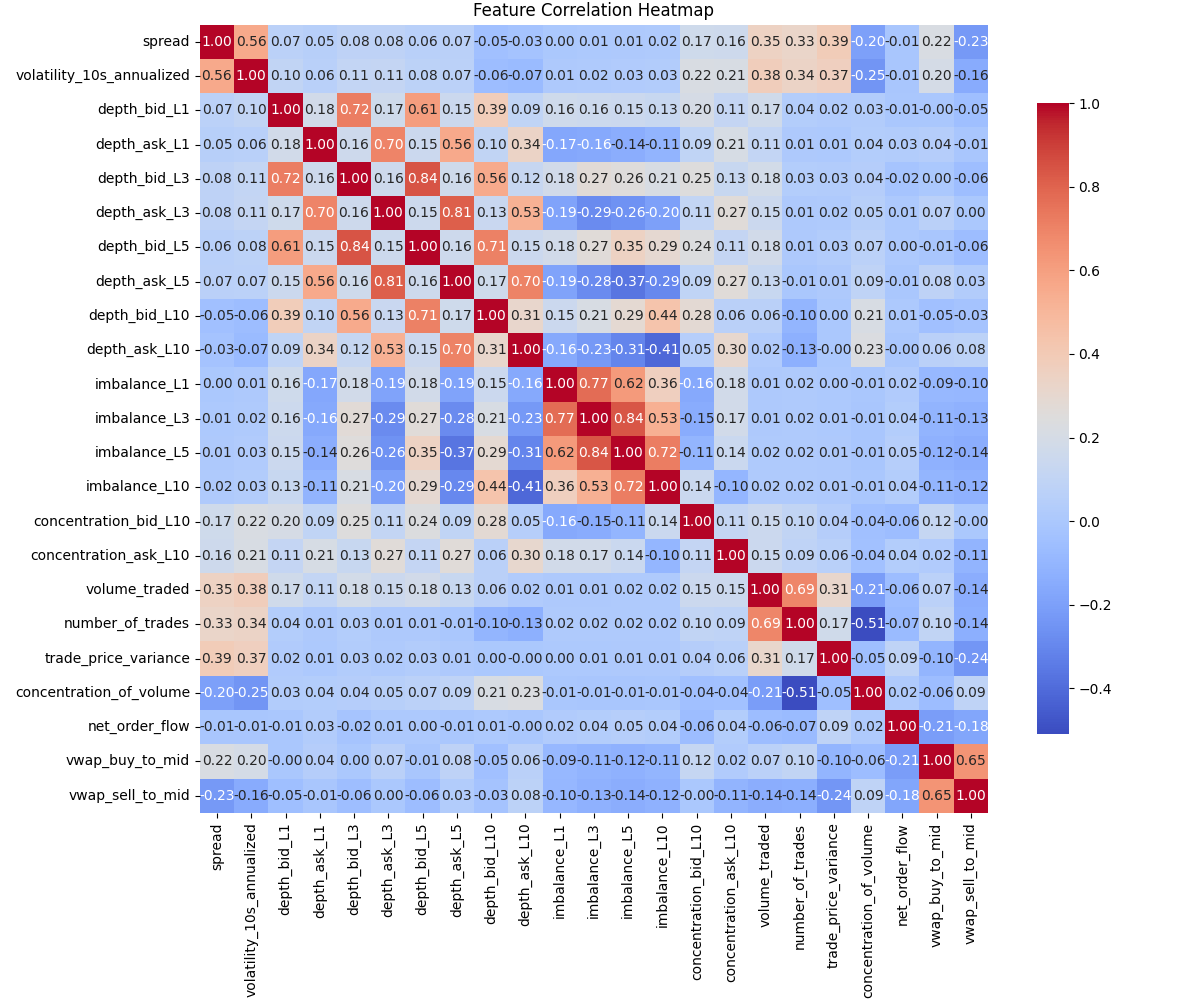}
    \caption{ETC}
  \end{subfigure}
  \begin{subfigure}{0.32\textwidth}
    \includegraphics[width=\linewidth]{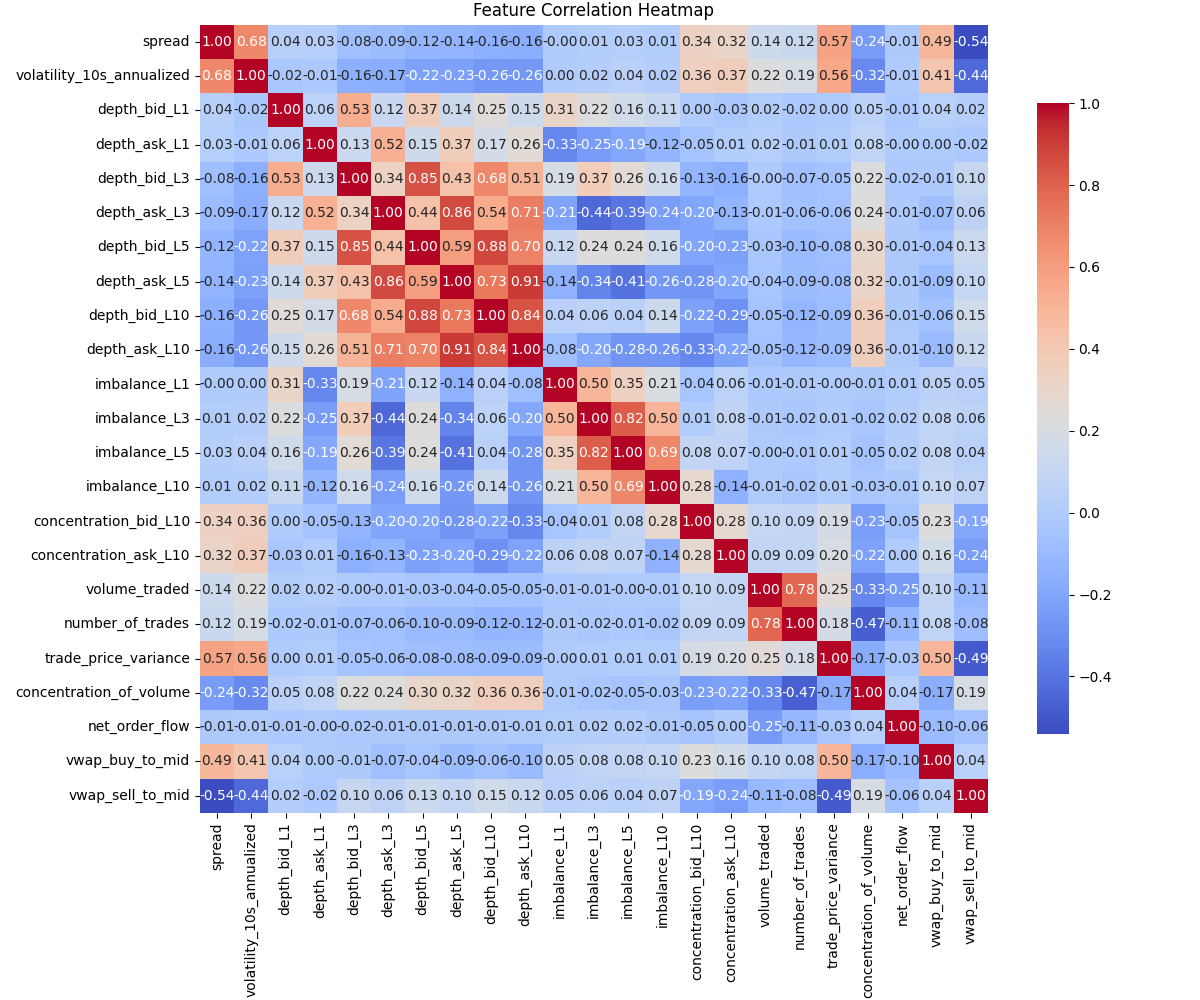}
    \caption{ROSE}
  \end{subfigure}
  \caption{The raw feature correlation heatmap shows some common patterns across assets, such as positive correlation in spreads and volatility, positive correlation in cumulative depths on the same side across various levels, or negative correlation in order book imbalance and VWAP-to-mid deviations. Correlation values remain quite similar across assets.}
\end{figure}

\section{Methodology}

For model evaluation, a rolling time-series cross-validation scheme is utilized. Each cross-validation fold is defined by training on a contiguous historical window and evaluating on subsequent, unseen periods, with a deliberate temporal gap between the training and validation sets. This temporal gap is critical to reduce information leakage, especially from features with slow time dynamics, ensuring that validation faithfully reflects genuine forward-looking predictive performance. The hyperparameters are tuned only within the training window using and inner time-series cross validation. The outer window is never used for tuning.

\begin{figure}[h]
    \centering
    \includegraphics[width=0.7\linewidth]{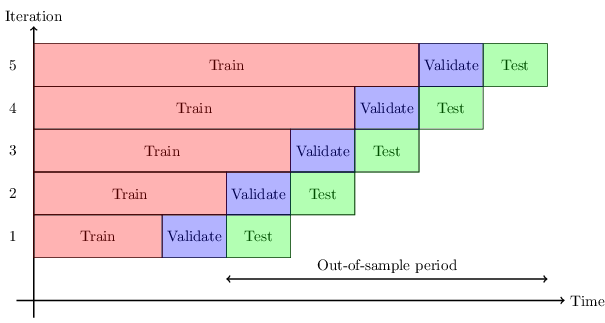}
    \caption{Schematic of walk-forward cross-validation with purging. In each fold, the model is trained on a rolling historical window and evaluated on a sequential future period, with a temporal gap (purge window) in between. This gap helps prevent information leakage by ensuring that slowly evolving features or delayed market responses do not contaminate the validation set, thus providing a more realistic estimate of true, forward-looking predictive performance.}
  \end{figure}

Two types of objective functions guide the training process. First, the standard squared error (and its associated out-of-fold \(R^2\)) serves as a baseline metric. More importantly for financial returns, we employ the Generalized Mean-Absolute Directional Loss (GMADL) as a metric. For a realized return \(R_i\) and corresponding prediction \(\hat{R}_i\), the GMADL is defined as:
\[
\ell_i = -\left(\frac{1}{1+e^{-a R_i \hat{R}_i}} - \frac{1}{2}\right) |R_i|^b, \qquad a, b > 0
\]
with the full objective being the mean over all samples. The GMADL explicitly emphasizes directional correctness, rewarding predictions that correctly match the sign of the realized return, and scales contributions by the magnitude of returns. This alignment is vital in return prediction tasks, where the practical utility of a forecast depends more on correct anticipation of large price moves than on minimizing squared error for small, noisy fluctuations. By using GMADL, the model is incentivized to focus on accuracy during periods of significant market action and to ignore trivial, low-amplitude fluctuations that might otherwise dominate squared error. GMADL is used as an evaluation metric, and models are trained with squared error loss funcion and selected by GMADL in-sample performance.

For each asset analyzed, we persist both the model trained to optimize the GMADL objective and the \(R^2\)-optimized variant. Unless otherwise specified, subsequent analyses and explanations are based on the GMADL model, with parallel checks on the alternative to confirm that results are not unduly sensitive to the choice of objective.

\subsection{Performance Metrics}

For each strategy and asset, several indicators are computed to evaluate
profitability and performance. When evaluating portfolio performance, it is critical to consider not just the return but also the risk of the strategy. In the study, we utilize performance metrics from Michańków et al. (2022) and Ryś and Ślepaczuk (2018).

\vspace{0.5cm}
\noindent
\textbf{Annualized Return Compounded}

\noindent
The Annualized Return Compounded (ARC), is the constant rate of annual return
over the whole period of investment, so that:

\begin{equation}
    ARC = \left(\frac{V(t_n)}{V(t_0)}\right)^{1/n} - 1
\end{equation}
where: \\
$V(t_0)$ - the initial value of the investment \\ 
$V(t_n)$ - the value of the investment at the end of the period \\ 
$n = t_n - t_0$ - number of years \\ 

\vspace{0.5cm}
\noindent
\textbf{Annualized Standard Deviation (ASD)}
\noindent
Volatility is a statistical indicator of the variation of returns. Most of the time, 
security is riskier the more volatile it is. Volatility may be expressed as either the standard
deviation or variation between returns from the same securities or market index. Volatility
might be easily switched to annualized values by multiplying the standard deviation of
the returns by the square root of the number of observations in a year (e.g. 252 for daily
data of the S\&P 500 Index and 365 for daily data of Bitcoin prices). In our research, we
use Annualized Standard Deviation (ASD) as a measure of volatility:

\begin{equation}
    ASD = \sqrt{\frac{1}{N-1}\sum_{t=1}^N (R_t - \hat{R})^2} \cdot \sqrt{n_{year}}
\end{equation}
where: \\
$\hat{R}$ - the average simple return (e.g. daily for daily data) of the given instrument \\ 
$R_t$ - the simple return during period t \\ 
$n_{year}$ -  number of observations in a year \\ 

\vspace{0.5cm}
\noindent
\textbf{Information Ratio}
\noindent
The Sharpe ratio, created by Nobel Prize winner William F. Sharpe, aids investors in
determining the return on investment relative to its risk. The ratio is the average return
over the risk-free rate (or some benchmark rate in the case of absolute return strategies)for each unit of volatility or overall risk. We assume a
zero-rate risk-free rate, therefore instead of Sharpe Ratio we will define this metric as IR:

\begin{equation}
    IR^* = \frac{ARC}{ASD}
\end{equation}

\vspace{0.5cm}
\noindent
\textbf{Max Drawdown (MDD)}
\noindent
A portfolio’s maximum drawdown (MDD) is the largest loss that could be recorded
between a portfolio’s peak and bottom before a new high is reached. Maximum drawdown serves as a gauge for the potential loss over a certain time frame. Maximum drawdown
(MDD), a major concern for most investors is a tool used to compare the relative riskiness
of different investment strategies.

\begin{equation}
    MDD(T) = \max_{t \in [0, T]}(\max_{t \in [0, T]}V_t - V_\tau)
\end{equation}

\vspace{0.5cm}
\noindent
\textbf{Max Loss Duration (MLD)}
\noindent
Maximum Loss Duration (MLD) is the worst (the greatest/longest) period
between peaks that the investment has experienced. It is expressed in the number of years:

\begin{equation}
    MLD = \max \frac{m_j - m_i}{S}
\end{equation}
\noindent
for which $Val(m_j) > Val(m_i)$ and $j > i. Val(m_j)$ and $Val(m_i)$ are the values of the local maximums in days $m_j$ and $m_i$ respectively. $m_j$ and mi are the numbers of days indicating
local maximums of the equity line. The scale parameter S denotes the number of trading
sessions in a year.

\vspace{0.5cm}
\noindent
\textbf{Information Ratio**}
\noindent
Kość et al. (2019) in their study use an additional measure to assess the effectiveness of
the strategy, which is a modification of the Information Ratio measure. This measure also
takes into account the sign of the portfolio’s rate of return and the maximum drawdown:

\begin{equation}
    IR^{**} = \frac{ARC^2 \cdot sign(ARC)}{ASD \cdot MDD}
\end{equation}

\section{Models}
\subsection{Gradient Boosted Decision Trees}

We employ gradient-boosted decision trees, specifically the CatBoost algorithm \citep{CatBoost2018}, to model the prediction target \(\rts\). CatBoost constructs an ensemble of shallow decision trees in a sequential manner, where each new tree corrects the residuals of the previous ensemble, allowing the model to flexibly capture nonlinear relationships, hierarchical effects, and complex interactions among features---patterns that are typical in market microstructure data. CatBoost is particularly well-suited for exploratory model fitting because it handles a wide range of heterogeneous, sometimes sparse or collinear tabular features without the need for extensive preprocessing, such as scaling or explicit encoding of categorical variables. Its built-in regularization and ordered boosting schemes help prevent overfitting, even in the presence of noisy or redundant predictors.

\subsection{Hyperparameter Optimization}

Hyperparameters critically govern the bias--variance trade-off, shaping the model's ability to generalize rather than overfit to idiosyncrasies in the training data. Key parameters include:
\begin{itemize}
    \item \textbf{Depth}: determines the interaction order between features.
    \item \textbf{Iterations and learning rate}: control functional complexity and capacity.
    \item \(\ell_2\) \textbf{leaf regularization} and \textbf{subsampling temperature}: prevent overfitting.
    \item \textbf{Discretization granularity}: balances data nuances against noise.
\end{itemize}

To navigate this high-dimensional search space efficiently, we employ automated hyperparameter optimization using the Optuna framework \citep{optuna}. Rather than relying on exhaustive grid search, Optuna implements a sophisticated Bayesian optimization algorithm (Tree-structured Parzen Estimator, TPE) to find optimal configurations. This process builds a probabilistic surrogate model of the objective function (e.g., out-of-fold validation loss) and uses an acquisition function to balance exploration and exploitation. We sample from broad, weakly informative prior ranges and select the configuration that generalizes best across our time-series cross-validation folds. We select the configuration using inner (training-only) time series cross validation. Performance is reported on outer folds only.

\subsection{Model Interpretation}

After model training, feature attribution is accomplished via SHAP (SHapley Additive exPlanations) \citep{LundbergLee2017}, using the TreeExplainer algorithm to compute values on held-out samples. SHAP provides several advantages compared to more traditional feature importance approaches. Unlike gain- or split-count-based measures that may be biased or local to a specific model realization, SHAP values are grounded in cooperative game theory and provide consistent, locally accurate attributions that account for feature interactions and correlations. This yields trustworthy global importance rankings and enables fine-grained analysis through feature dependence plots, which emulate partial dependence while retaining the conditional distributions and dependencies learned by the models.

We monitor out-of-fold \(R^2\) and the direction-aware objective, and we examine SHAP summaries for logically inconsistent patterns such as dominance by a single proxy or inconsistent signs across folds. In practice, the same few features repeatedly surface as most influential across assets, and their SHAP signs and shapes agree with microstructure priors. For each asset, we carry forward the model that optimizes the GMADL objective for explanation and backtesting, while also retaining the \(R^2\)-optimized model as a robustness check.

\section{SHAP Explanations}
We compute SHAP values on held-out samples using TreeExplainer and analyze both global and local structure. Across assets, summary plots consistently highlight a common set of features: order flow imbalance, bid–ask spreads, and VWAP-to-mid deviations dominate mean absolute SHAP. This suggests that the same economic channels drive short-horizon returns irrespective of capitalization, and the stability persists when replacing the training objective with squared error or modestly shifting the prediction horizon.

\begin{figure}[h]
  \centering
  \begin{minipage}{0.32\textwidth}
    \centering
    \includegraphics[width=\linewidth]{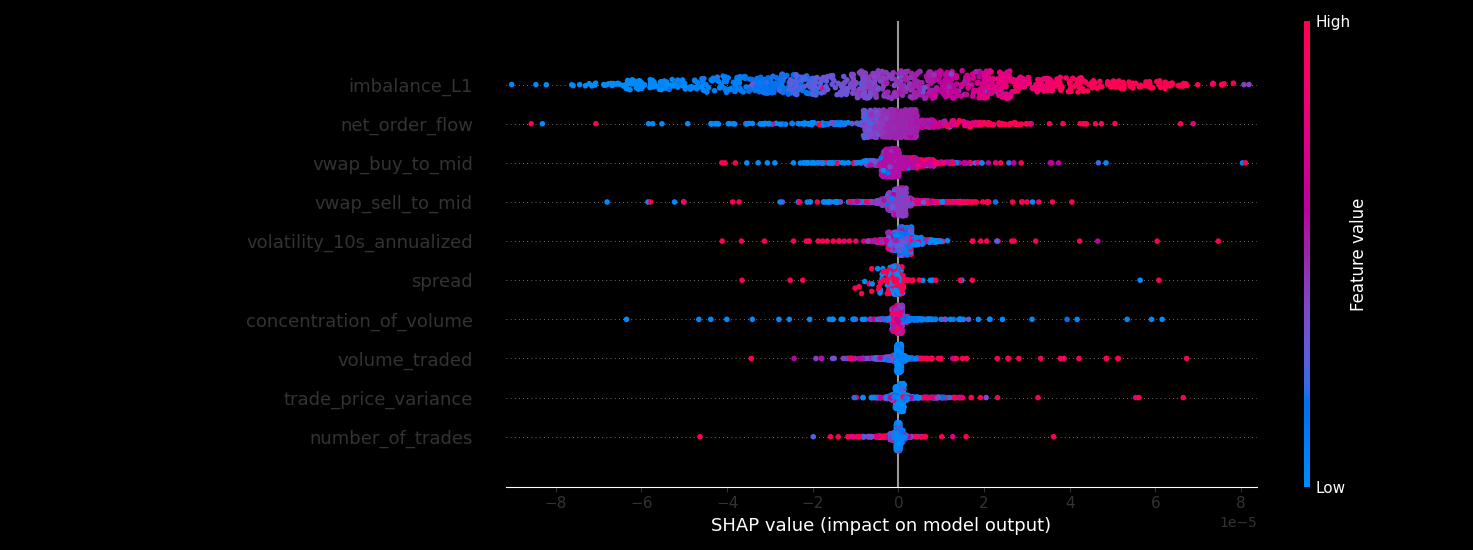}
    \vspace{2mm}
    \textbf{BTC}
  \end{minipage}%
  \hfill
  \begin{minipage}{0.32\textwidth}
    \centering
    \includegraphics[width=\linewidth]{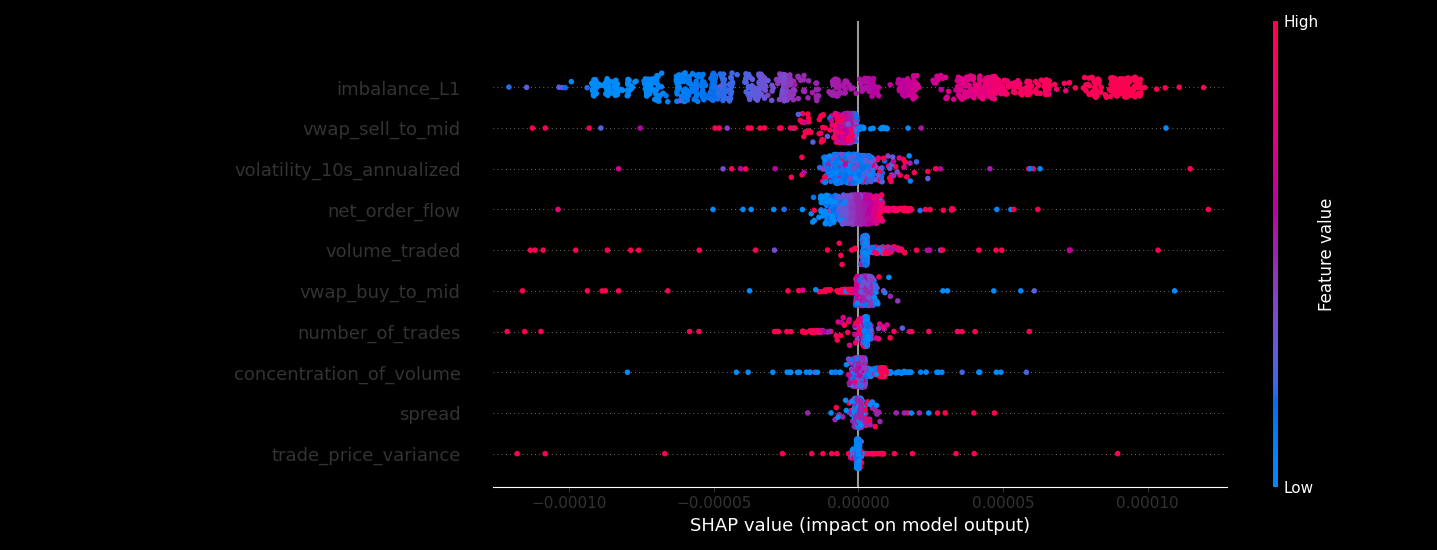}
    \vspace{2mm}
    \textbf{LTC}
  \end{minipage}%
  \hfill
  \begin{minipage}{0.32\textwidth}
    \centering
    \includegraphics[width=\linewidth]{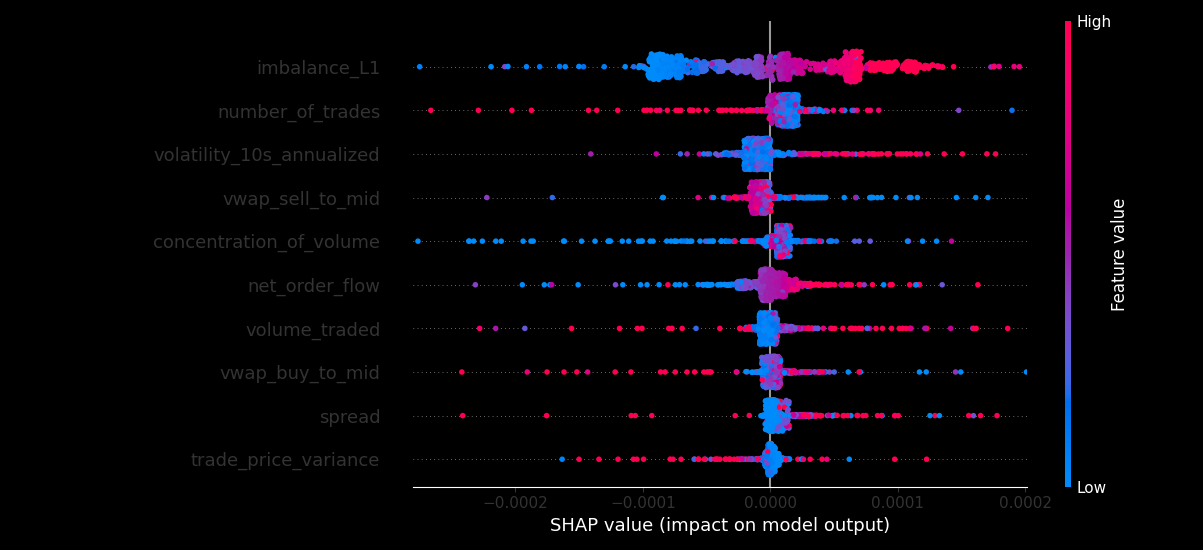}
    \vspace{2mm}
    \textbf{ETC}
  \end{minipage}
  \caption{Global SHAP summaries for BTC, LTC, and ETC shown side by side.}
\end{figure}

Let \(S_j\) denote the mean absolute SHAP for feature \(j\) within a given asset. The rankings induced by \(S_j\) are highly correlated across assets, indicating that variables transmitting liquidity-taking pressure and adverse selection uniformly carry the largest marginal contributions. Beyond rankings, the SHAP dependence curves exhibit consistent shapes across assets: the effect of order flow imbalance on returns is predominantly monotone with concavity at extremes (diminishing incremental impact as pressure accumulates); wider spreads associate with attenuated predictive effects and lower-confidence signals, in line with elevated adverse selection risk and execution costs; and VWAP-to-mid deviations display short-horizon asymmetries coherent with transient pressure followed by microstructure reversion as depth replenishes.



\begin{figure}[h]
  \centering
  \begin{minipage}{0.31\textwidth}
    \includegraphics[width=\linewidth]{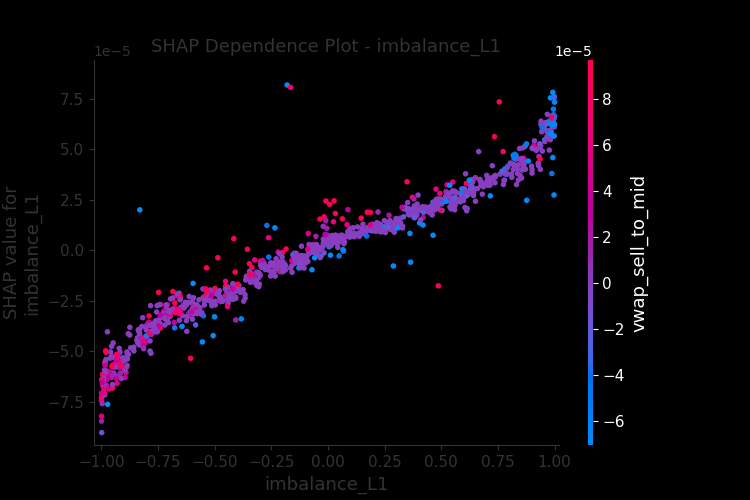}
    \vspace{2mm}
    \centering
    \textbf{BTC}
  \end{minipage}
  \hfill
  \begin{minipage}{0.31\textwidth}
    \includegraphics[width=\linewidth]{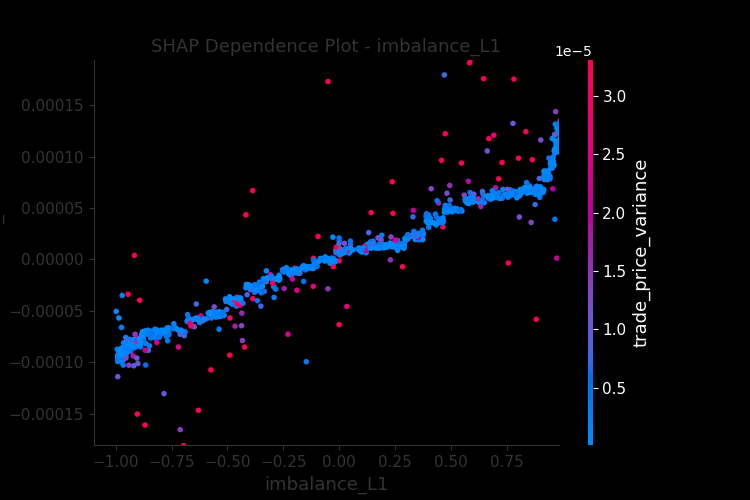}
    \vspace{2mm}
    \centering
    \textbf{ETC}
  \end{minipage}
  \hfill
  \begin{minipage}{0.31\textwidth}
    \includegraphics[width=\linewidth]{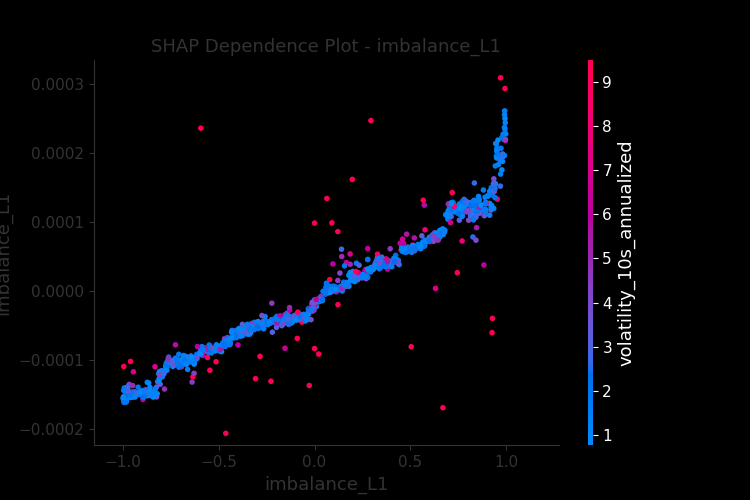}
    \vspace{2mm}
    \centering
    \textbf{ROSE}
  \end{minipage}
  \caption{SHAP dependence plots (top feature: orderbook imbalance) for BTC, ETC, and ROSE shown side by side.}
\end{figure}

Overlaying dependence plots reveals similar slopes near the origin and comparable saturation levels across assets with very different tick sizes and liquidity, pointing to scale-free structure once expressed in relative prices and flows.

\subsection{Note on the relation between imbalance effect magnitude and tick size}
Tick size modulates how order flow pressure translates into discrete price moves and thus shapes the predictive power of top-of-book imbalance. Evidence across assets shows that the predictive effect of imbalance (as captured by SHAP) is stronger for instruments with larger ticks. This is consistent with the microprice mechanism: when ticks are large relative to the local price scale, the probability of an uptick (or downtick) inferred from depth asymmetry maps more directly into the next price change, amplifying the marginal contribution of imbalance features.

\begin{figure}[h]
  \centering
  \includegraphics[width=0.7\linewidth]{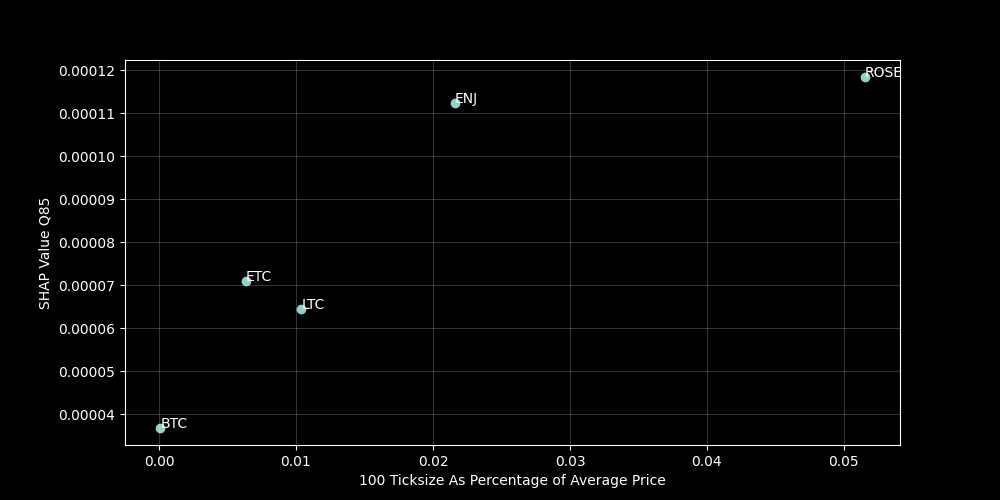}
  \caption{Association between relative tick size and high-quantile imbalance SHAP values across assets. Larger relative ticks correspond to stronger imbalance contributions.}
\end{figure}

The cross-asset relationship between an estimate of effective tick size (scaled by the average price) and an upper-quantile summary of imbalance SHAP values displays a positive association: assets with larger relative ticks exhibit higher high-quantile SHAP magnitudes for imbalance. This aligns with a discrete-move view of impact and with inventory absorption limits that bind more readily under coarse price grids.

\begin{figure}[H]
  \centering
  \includegraphics[width=0.8\linewidth]{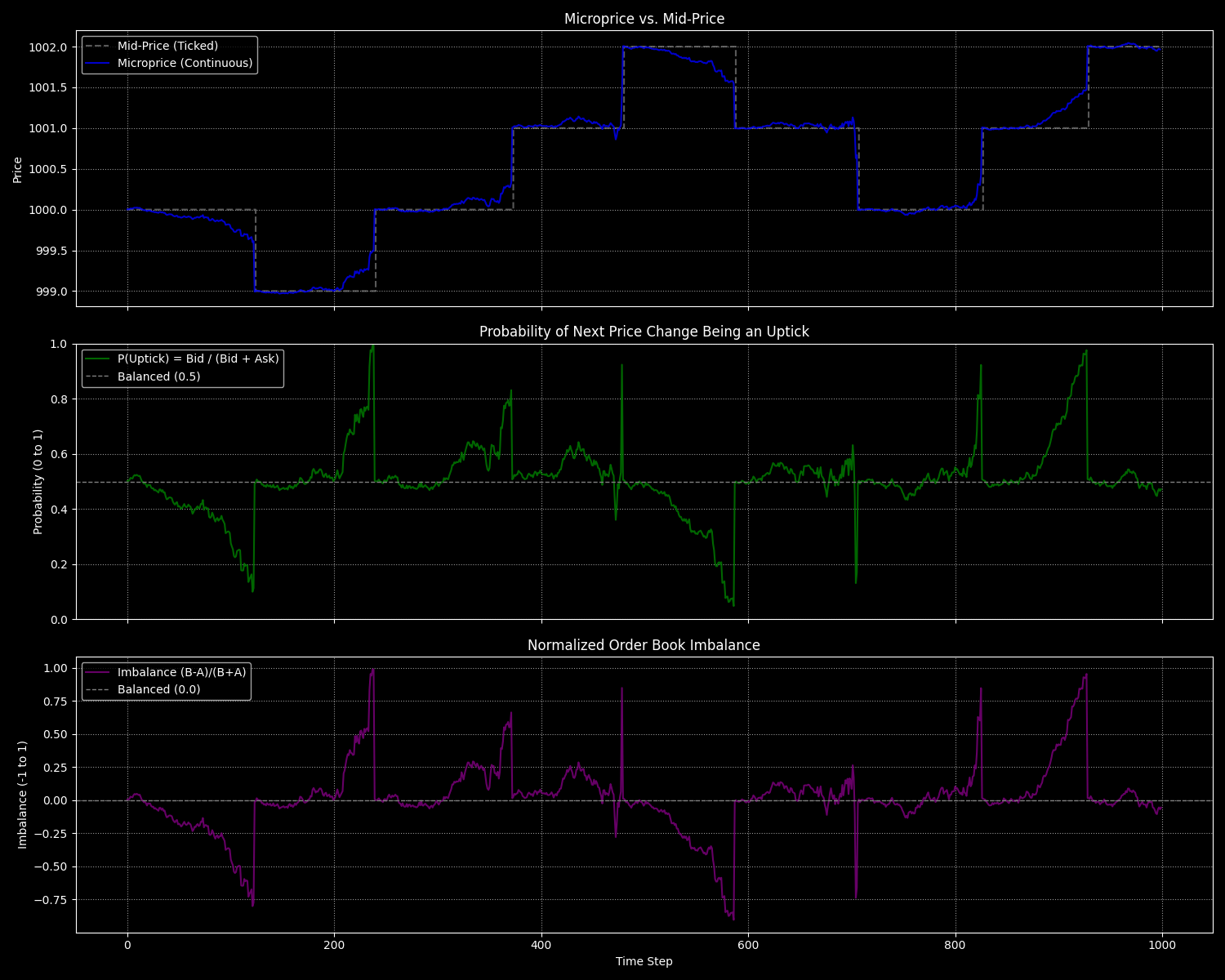}
  \caption{Microprice versus ticked mid in a depth-replenishment simulation. Imbalance-induced shifts in microprice anticipate discrete mid changes, with stronger effects under coarser ticks.}
\end{figure}

A stylized simulation with replenishing depth illustrates the link between depth asymmetry, the probability of an uptick, and a continuous microprice relative to a ticked mid. The microprice tracks the ticked mid with anticipatory shifts proportional to the imbalance-implied uptick probability, clarifying why coarser ticks strengthen the mapping from imbalance to realized returns at short horizons.

\begin{figure}[H]
  \centering
  \includegraphics[width=0.8\linewidth]{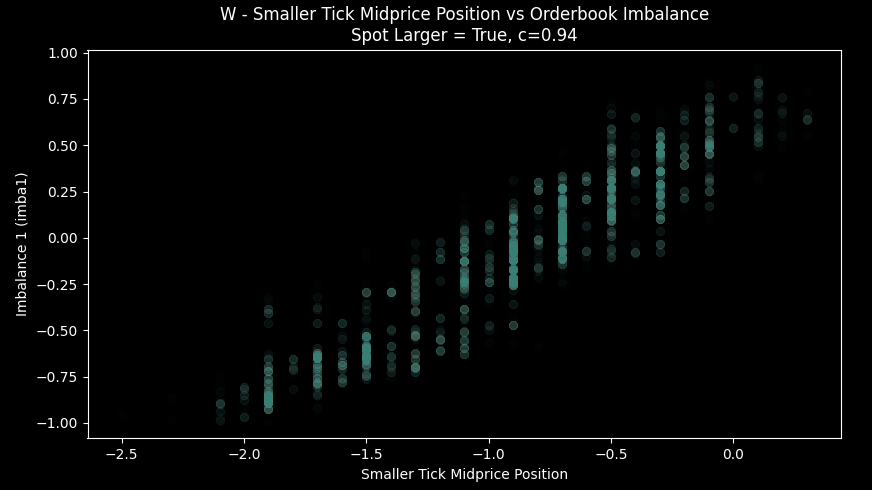}
  \caption{Scatter plot of the smaller-tick futures mid-price position within the larger-tick spot spread  $(2 * (mid_{fut} - bid_{spot}) / (ask_{spot} - bid_{spot}) - 1)$ versus the spot order book imbalance for asset W. The strong positive correlation ($c=0.94$) indicates that spot imbalance accurately reflects the continuous price location.}
  \label{fig:small_midprice_pos}
\end{figure}

This relationship is further validated by comparing the order book imbalance of a large-tick spot instrument (W/USDT) with the price movements of its corresponding futures contract (W/USDT Perpetual), which has a significantly finer tick size ($10^{-5}$ vs $10^{-4}$). As shown in Figure \ref{fig:small_midprice_pos}, the futures mid-price (which serves as a proxy for the unobserved continuous efficient price) exhibits variations within the wider spot spread that are highly correlated with the spot order book imbalance ($c=0.94$). When the spot book is heavy on the bid side (positive imbalance), the futures price tends to trade near the upper bound of the spot bid-ask spread, and vice versa. This empirical evidence confirms that order book imbalance in large-tick markets effectively acts as a visible proxy for the latent continuous price location within the spread.

\begin{figure}[H]
  \centering
  \includegraphics[width=0.8\linewidth]{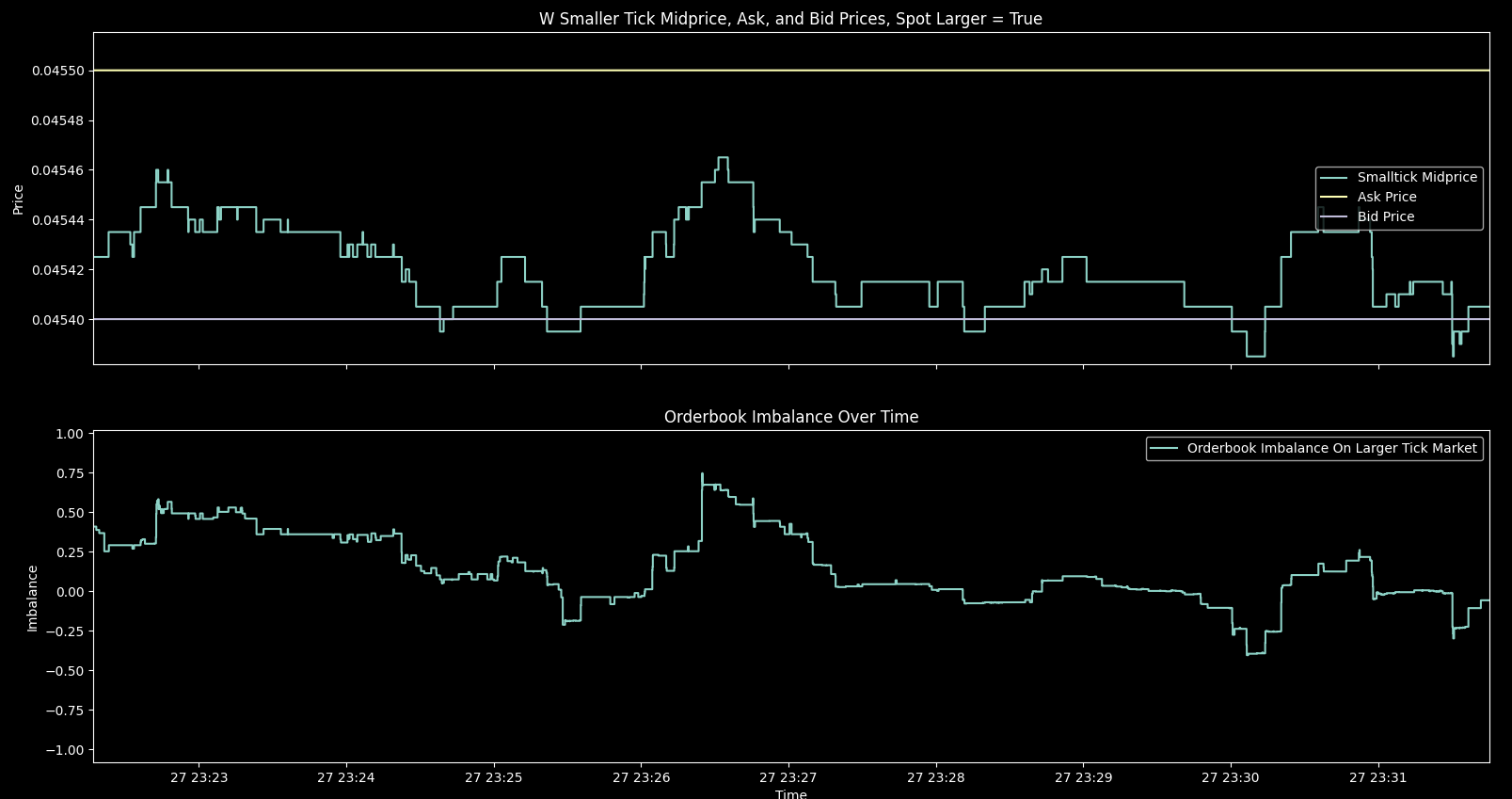}
  \caption{Plot comparing futures position within spot orderbook spread (top) versus spot orderbook imbalance (bottom) for asset W.}
  \label{fig:futprice_vs_imba}
\end{figure}

\begin{figure}[H]
  \centering
  \includegraphics[width=0.5\linewidth]{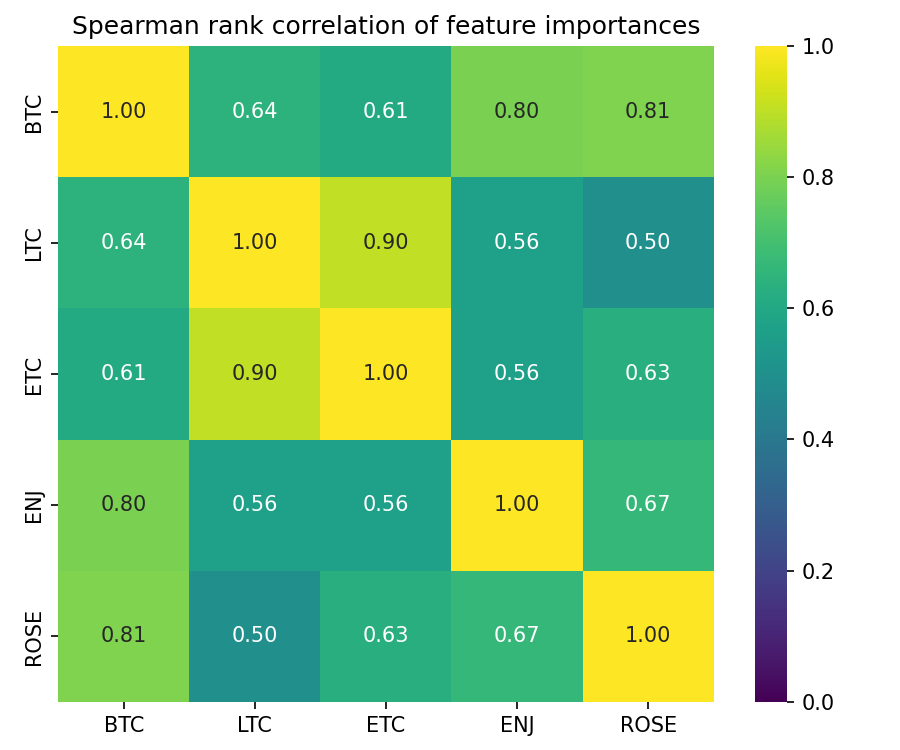}
  \caption{Spearman rank correlation of feature importances (mean absolute SHAP) across assets. High correlation indicates that the most important predictive features are consistent across different cryptocurrencies.}
\end{figure}

\section{Backtesting Strategies}
\subsection{Taker Execution}

We assess economic significance using a conservative taker execution framework (submitting market orders to buy or sell based on the model signal) designed to understate performance. Signals are derived from model outputs by applying a symmetric threshold \(\theta\) and taking the sign of the prediction when \(|\hat R_t|>\theta\). The prediction target is a 3-second mid-price return, while position managment is even driven, therefore holding times can be shorter than 3 seconds.Positions are adjusted only on signal changes to avoid spurious turnover.

Buys are executed at the best ask and sells at the best bid, reflecting realistic taker costs. Inventory is marked to the unfavorable side of the book (longs to the bid, shorts to the ask) so that unrealized PnL is systematically penalized relative to mid-marking. Cash accounts are updated at each trade; equity equals cash plus marked inventory value. This construction ensures that any reported PnL is not an artifact of optimistic pricing.

We use fixed notional trades to isolate signal quality from leverage choices. Optional proportional costs can be introduced to reflect taker fees; we report gross and fee adjusted returns using taker fee. Latency is not explicitly modeled; results should be interpreted as an upper bound in the fastest regime.

We summarize results by metrics explained in the methodology section. Equity curves provide a visual diagnostic of signal stability and resilience to brief regime shifts. Thresholds \(\theta\) are selected to balance participation and confidence; higher thresholds concentrate trades on high-magnitude forecasts with improved precision but reduced activity.

\begin{figure}[H]
    \centering
    \includegraphics[width=0.9\textwidth]{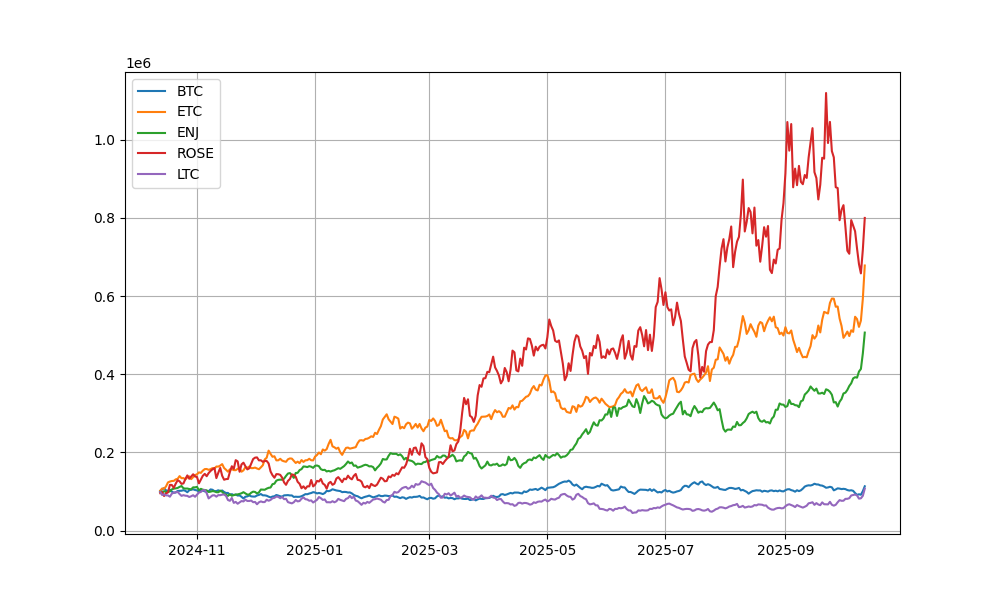}
    \caption{Equity curve for the taker backtest.}
    \label{fig:taker_equity}
\end{figure}

\begin{table}[H]
    \centering
    \small
    \rule{\textwidth}{0.4pt}
    \begin{tabular}{lrrrrrrrrrr}
\toprule
asset & btc & btc-bh & enj & enj-bh & etc & etc-bh & ltc & ltc-bh & rose & rose-bh \\
\midrule
ARC & 0.13 & 0.83 & 4.06 & -0.67 & 5.78 & -0.10 & 0.07 & 0.53 & 7.00 & -0.71 \\
ASD & 0.53 & 0.45 & 0.62 & 1.13 & 0.64 & 0.85 & 0.99 & 0.86 & 1.33 & 1.10 \\
IR* & 0.25 & 1.87 & 6.58 & -0.59 & 8.97 & -0.12 & 0.07 & 0.62 & 5.28 & -0.64 \\
MDD & 0.29 & 0.28 & 0.26 & 0.89 & 0.24 & 0.63 & 0.64 & 0.50 & 0.43 & 0.87 \\
MLD-years & 0.53 & 0.32 & 0.22 & 0.85 & 0.20 & 0.85 & 0.63 & 0.73 & 0.21 & 0.85 \\
IR** & 0.12 & 5.53 & 101.21 & -0.44 & 213.58 & -0.02 & 0.01 & 0.66 & 86.02 & -0.52 \\
\end{tabular}

    \rule{\textwidth}{0.4pt}
    \caption{Rounded summary statistics for the taker backtest across assets. ARC = Annualized return on capital, ASD = annualized standard deviation, IR = information ratio, MDD = max drawdown, MLDyears = max loss duration (years)}
    \label{tab:taker_metrics_manual}
\end{table}

\subsection{Maker Execution}

We also evaluate a market-maker framework using identical signals but maker-style execution. Positions are entered passively by posting limit orders at the bid (for buys) or ask (for sells), with fills simulated based on subsequent trade-through and queue priority models. As in the taker test, equity is marked pessimistically (long inventory to bid, short to ask) to avoid overstatement from optimistic mid-point assumptions.

Maker PnL thus captures revenue from spread capture and signal efficacy, net of adverse selection and missed fills. Costs include potential adverse selection, but not taker fees, making results directly comparable to standard market making. Fixed notional position sizing is used for comparability.

\begin{figure}[h!]
    \centering
    \includegraphics[width=0.9\textwidth]{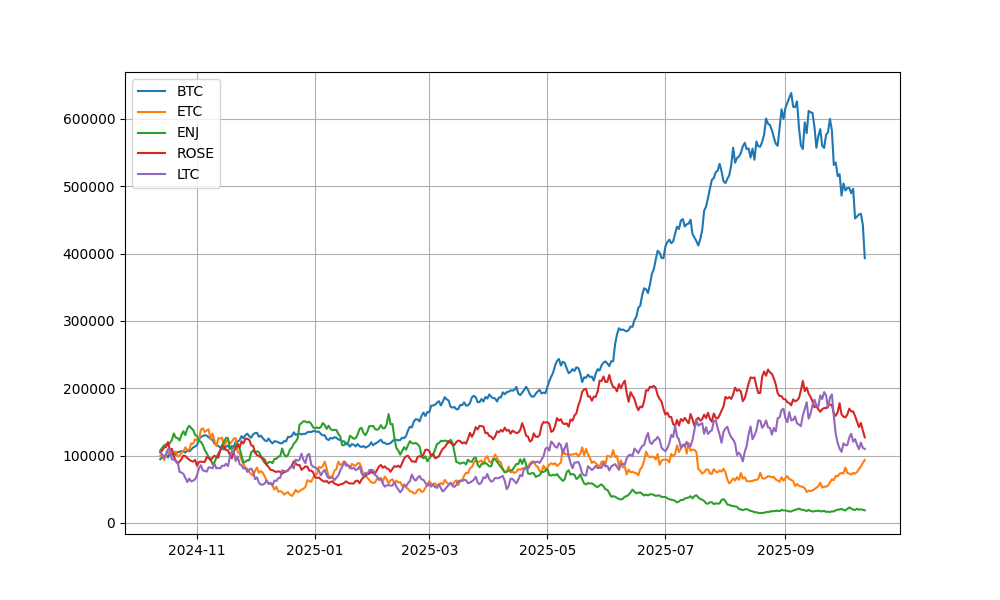}
    \caption{Equity curve for the maker backtest.}
    \label{fig:maker_equity}
\end{figure}

\begin{table}[h!]
    \centering
    \small
    \rule{\textwidth}{0.4pt}
    \begin{tabular}{lrrrrrrrrrr}
\toprule
asset & btc & btc-bh & enj & enj-bh & etc & etc-bh & ltc & ltc-bh & rose & rose-bh \\
\midrule
ARC & 2.93 & 0.83 & -0.81 & -0.67 & -0.07 & -0.10 & 0.10 & 0.53 & 0.27 & -0.71 \\
ASD & 0.54 & 0.45 & 1.05 & 1.13 & 1.31 & 0.85 & 1.40 & 0.86 & 0.84 & 1.10 \\
IR* & 5.47 & 1.87 & -0.77 & -0.59 & -0.05 & -0.12 & 0.07 & 0.62 & 0.32 & -0.64 \\
MDD & 0.38 & 0.28 & 0.91 & 0.89 & 0.71 & 0.63 & 0.59 & 0.50 & 0.55 & 0.87 \\
MLD-years & 0.13 & 0.32 & 0.67 & 0.85 & 0.94 & 0.85 & 0.53 & 0.73 & 0.31 & 0.85 \\
IR** & 41.78 & 5.53 & -0.69 & -0.44 & -0.00 & -0.02 & 0.01 & 0.66 & 0.16 & -0.52 \\
\end{tabular}

    \rule{\textwidth}{0.4pt}
    \caption{Rounded summary statistics for the maker backtest across assets. ARC = Annualized return on capital, ASD = annualized standard deviation, IR = information ratio, MDD = max drawdown, MLDyears = max loss duration (years)}
    \label{tab:maker_metrics_manual}
\end{table}

\subsection{Combined Strategy and Discussion}

\begin{figure}[h!]
    \centering
    \includegraphics[width=0.9\textwidth]{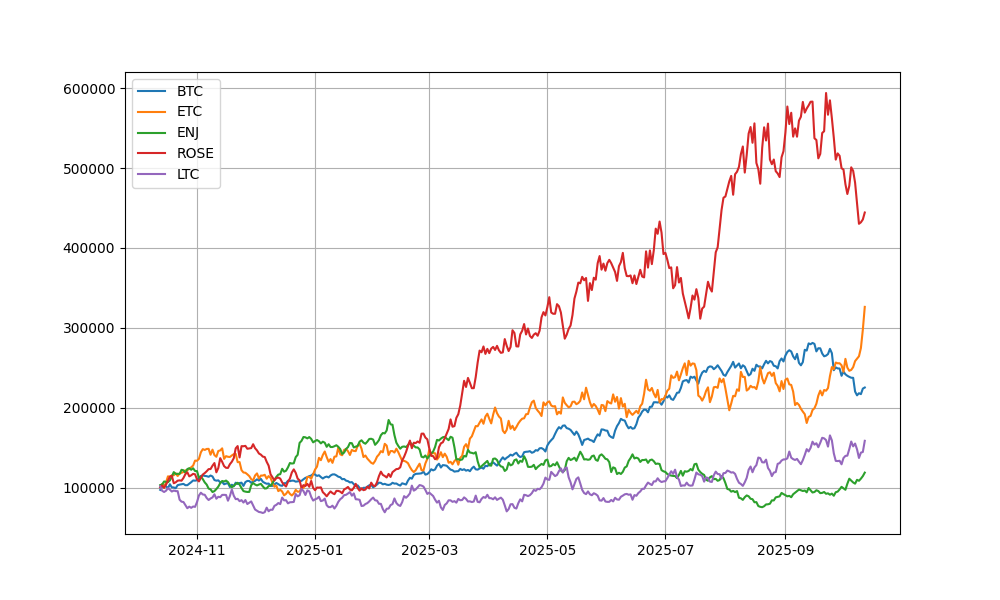}
    \caption{Equity curve for the averaged backtest.}
    \label{fig:avg_equity}
\end{figure}

\begin{table}[h!]
    \centering
    \small
    \begin{tabular}{lrrrrrrrrrr}
\toprule
asset & btc & btc-bh & enj & enj-bh & etc & etc-bh & ltc & ltc-bh & rose & rose-bh \\
\midrule
ARC & 1.25 & 0.83 & 0.19 & -0.67 & 2.26 & -0.10 & 0.59 & 0.53 & 3.44 & -0.71 \\
ASD & 0.39 & 0.45 & 0.59 & 1.13 & 0.74 & 0.85 & 0.84 & 0.86 & 0.75 & 1.10 \\
IR* & 3.24 & 1.87 & 0.32 & -0.59 & 3.05 & -0.12 & 0.70 & 0.62 & 4.60 & -0.64 \\
MDD & 0.23 & 0.28 & 0.59 & 0.89 & 0.40 & 0.63 & 0.34 & 0.50 & 0.42 & 0.87 \\
MLD-years & 0.15 & 0.32 & 0.67 & 0.85 & 0.22 & 0.85 & 0.35 & 0.73 & 0.22 & 0.85 \\
IR** & 17.43 & 5.53 & 0.10 & -0.44 & 17.47 & -0.02 & 1.19 & 0.66 & 37.52 & -0.52 \\
\midrule
\end{tabular}

    \caption{Rounded summary statistics for the averaged backtest across assets. ARC = Annualized return on capital, ASD = annualized standard deviation, IR = information ratio, MDD = max drawdown, MLDyears = max loss duration (years)}
    \label{tab:joined_metrics_manual}
\end{table}

We additionally evaluated a blended strategy allocating 50\% of capital to the taker execution and 50\% to the maker execution. While the strategies exhibit negative correlation during extreme events like the October 10th flash crash, they remain positively correlated during normal regimes as they rely on identical predictive features. Consequently, although the equal-weighted combination reduces some volatility, the generally weaker performance of the maker component tends to drag down the aggregate results compared to the pure taker approach. This reinforces the finding that, for these specific predictive signals, the information advantage is most effectively monetized through immediate liquidity taking rather than passive provision, particularly given the adverse selection risks inherent in simple limit order strategies.

Overall, both taker and maker strategies exhibit profitability across most assets, but there are notable differences in their performance profiles. The maker strategy generally achieves higher annualized returns on capital (ARC) and information ratios (IR) for the top coin (BTC) and respectable performance on others, though with more modest results or even slight losses on less liquid assets like LTC. The taker strategy, while profitable in aggregate, shows a particularly outsized return for ROSE, largely due to the extreme event on 2025-10-10.

\begin{table}[h!]
    \centering
    \footnotesize
    \begin{tabular}{lrrrrrr}
\toprule
 & Taker t-stat & Taker p-value & Maker t-stat & Maker p-value & Avg t-stat & Avg p-value \\
instrument &  &  &  &  &  &  \\
\midrule
BTC & -0.6671 & 0.7474 & 1.2523 & 0.1056 & 0.3724 & 0.3549 \\
ETC & \textbf{1.7208} & \textbf{0.0431} & 0.3406 & 0.3668 & 1.0368 & 0.1503 \\
ENJ & \textbf{1.7942} & \textbf{0.0368} & -0.4677 & 0.6799 & 0.5830 & 0.2801 \\
ROSE & \textbf{2.0777} & \textbf{0.0192} & 0.8045 & 0.2108 & \textbf{1.7440} & \textbf{0.0410} \\
LTC & -0.1736 & 0.5688 & 0.1392 & 0.4447 & 0.0039 & 0.4985 \\
\midrule
\end{tabular}

    \caption{T-test results for the taker and maker backtest across assets (results statistically significant at $5\%$ level are bolded).}
    \label{tab:ttest_table}
\end{table}

The t-test table (\autoref{tab:ttest_table}) shows which strategies have mean returns statistically different from buy and hold mean return. For taker strategies, the p-value is below $0.05$ for ETC, ENJ, and ROSE, indicating statistical significance, while for BTC and LTC, the results are not significant. For maker strategies, all p-values exceed $0.05$, so the null hypothesis of zero mean returns cannot be rejected for any asset in the maker backtest. Thus, only the taker strategies on ETC, ENJ, and ROSE demonstrate statistically significant outperformance at the $5\%$ level.

This disparity during the flash crash can be understood in microstructure terms. In such rapid price dislocation, taker strategies, able to execute instantly at available prices, can capitalize on large directional moves and exploit temporary liquidity gaps, thus achieving windfall profits if the model correctly predicts the direction. In contrast, makers provide passive liquidity and are exposed to adverse selection: during abrupt crashes, their outstanding limit orders (especially stale bids) may be aggressively lifted, leading to significant losses before they can adjust quotes or inventory. This vulnerability to "picking off" risk explains the observed drawdown for makers during extreme events, whereas takers benefit from flexibility and speed at the cost of paying the spread and fees in normal conditions.

Therefore, while maker strategies generally offer steady spread capture and lower risk under ordinary conditions, they bear heightened exposure during disorderly markets. Taker strategies, conversely, are more robust to regime shifts but may incur higher steady-state costs. The specific outcomes observed highlight this fundamental microstructural distinction.

\section{Performance During October 10th Flash Crash}
\label{sec:robustness}

October 2025 was anticipated by many market participants to be another "Uptober," a month characterized by seasonal bullish price action in the cryptocurrency markets. Instead, it became the setting for one of the most severe market dislocations in recent history. On October 10, a surprise announcement of significant tariffs on Chinese imports by the U.S. administration triggered a catastrophic flash crash. The event led to the largest single-day wipeout in crypto history, with over \$19 billion in leveraged positions liquidated within 24 hours. Bitcoin (BTC) plummeted 18\% from its all-time high, and the broader altcoin market suffered catastrophic losses as liquidity evaporated and sentiment shifted to "Extreme Fear." \thanks{https://www.fticonsulting.com/insights/articles/crypto-crash-october-2025-leverage-met-liquidity}

This "black swan" event, a full-scale deleveraging that reset market dynamics, provides a perfect, albeit brutal, natural experiment for our trading model. A crucial test for any trading strategy is its performance during such periods of extreme market stress. While profitability in calm, well-behaved markets is desirable, a strategy's true robustness is revealed by its behavior during unexpected events. The October 10 flash crash serves as an acid test, exposing vulnerabilities and validating the core predictive power of our model. In this section, we analyze the performance of our taker strategy during this event to assess its resilience and explore its broader implications for market stability.

\subsection{Baseline Performance in Quiescent Markets}

\begin{figure}[h!]
    \centering
    \includegraphics[width=\textwidth]{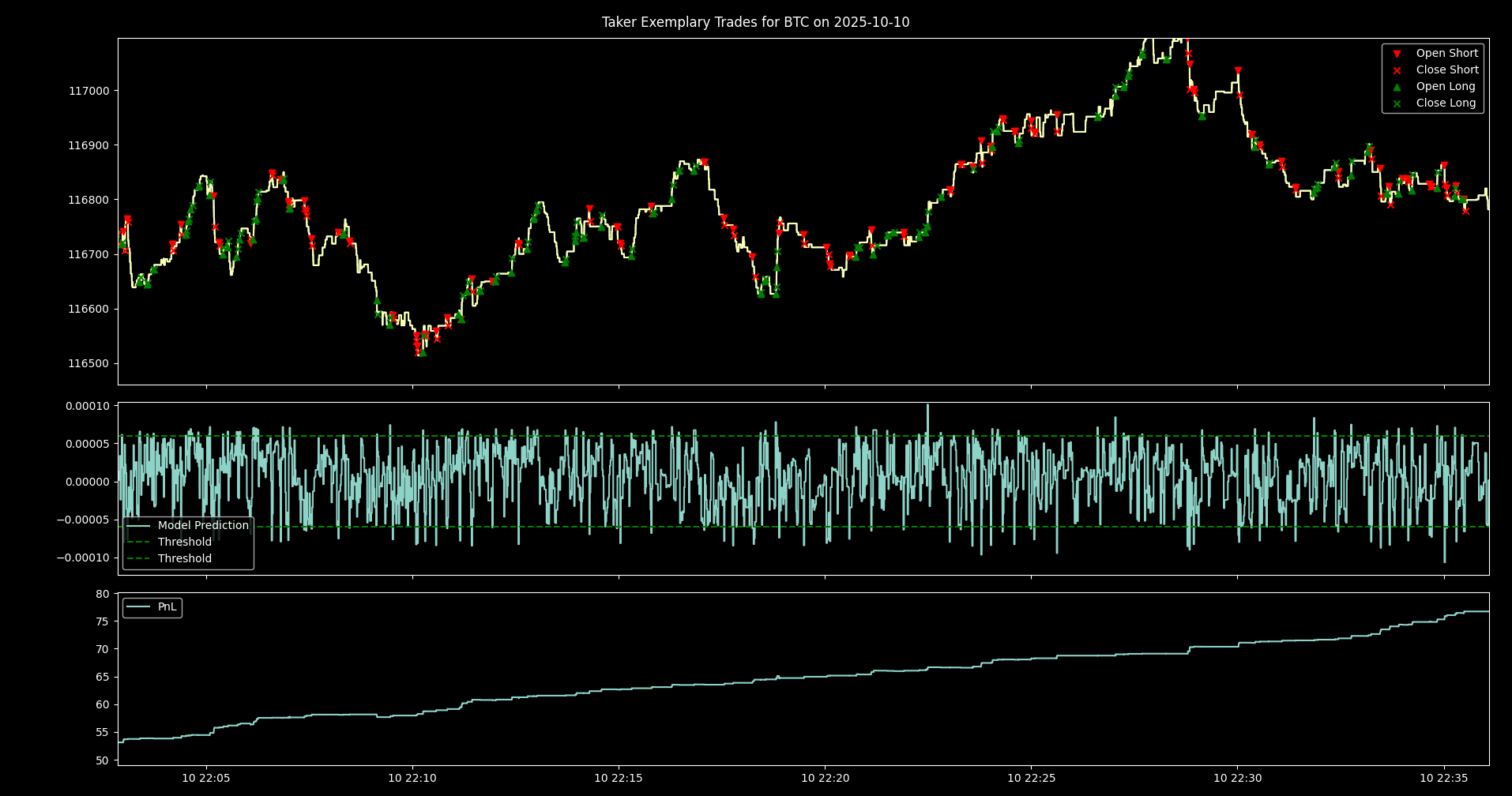}
    \caption{Taker strategy performance during a period of normal market activity. The top panel shows prices and trades, the middle panel displays model predictions and thresholds, and the bottom panel shows the cumulative PnL.}
    \label{fig:taker_ordinary}
\end{figure}

Under normal market conditions, our taker strategy performs as a high-frequency algorithm, designed to capture small, fleeting price discrepancies. Figure \ref{fig:taker_ordinary} illustrates this typical behavior. The model's predictions remain stationary, oscillating within a tight band around zero, triggering frequent, short-lived trades that typically last for only one to two seconds. The corresponding PnL curve exhibits a steady, gradual increase, reflecting the accumulation of small profits from a large number of trades.

This baseline performance demonstrates the model's effectiveness in identifying subtle, mean-reverting patterns in the order book imbalance. However, it is crucial to acknowledge the limitations of a backtesting environment. Real-world execution is subject to latency, network jitter, and queue position dynamics, which are notoriously difficult to simulate accurately. The profits observed in this idealized setting may not be fully realizable in a live trading environment where microsecond advantages are paramount.

\subsection{Performance during the October 10, 2025 Flash Crash}

\begin{figure}[h!]
    \centering
    \includegraphics[width=\textwidth]{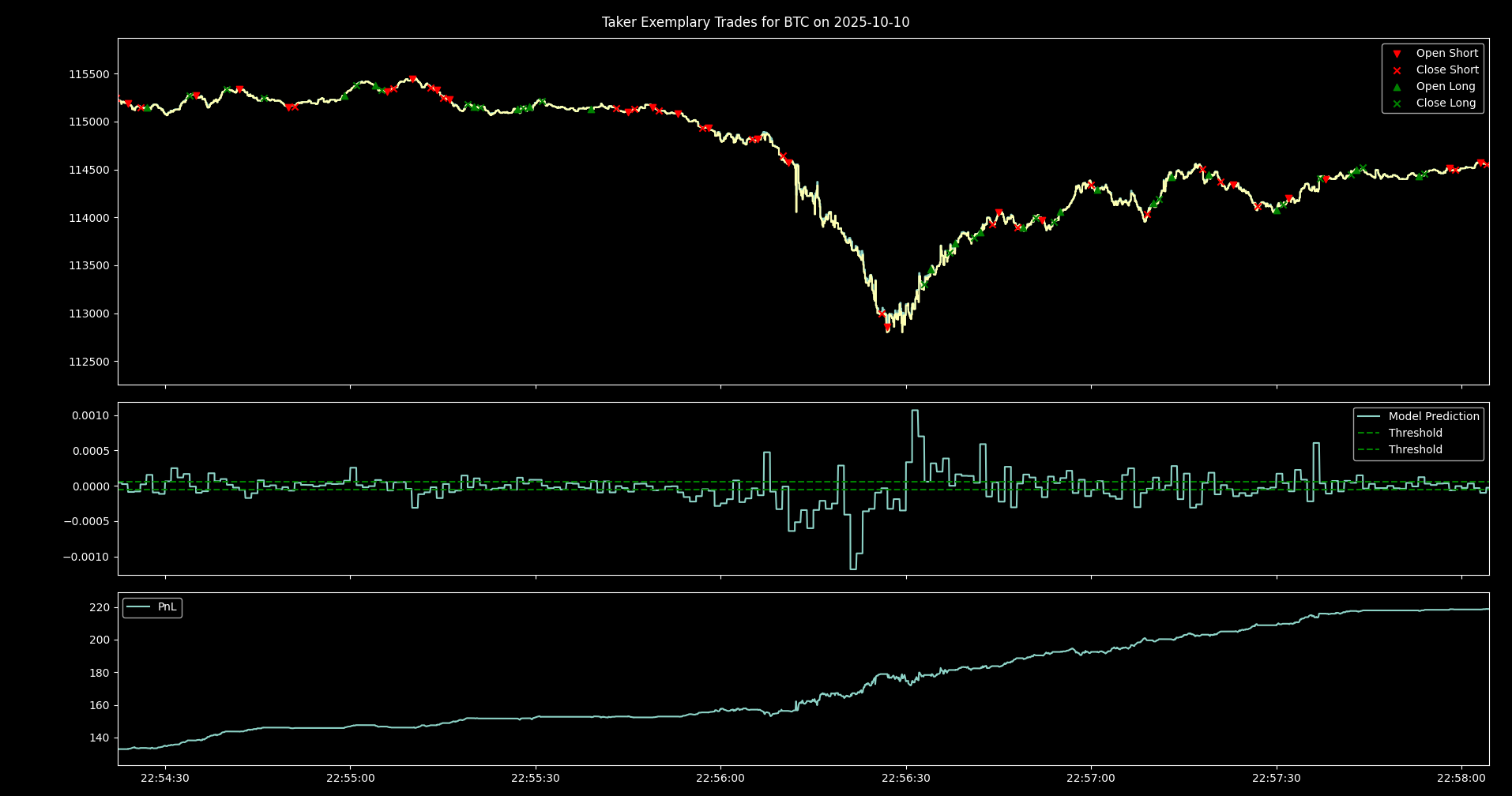}
    \caption{Taker strategy performance during the October 10, 2025 flash crash. The model correctly enters a short position, capturing a significant profit from the price decline.}
    \label{fig:taker_fc1}
\end{figure}

The flash crash on October 10, 2025, provided a natural experiment to test our model under extreme conditions. As depicted in Figure \ref{fig:taker_fc1}, the event was characterized by a sudden and dramatic price collapse in Bitcoin. Our model, which primarily relies on order book imbalance, successfully detected the build-up of selling pressure and initiated a short position. This timely action resulted in a significant profit, validating the paper's core thesis that order book imbalance is a potent predictor of near-term price movements, especially during periods of market instability.

Notably, the model's behavior deviated significantly from its baseline. The holding period for the short position was an outlier, lasting approximately 20 seconds---an order of magnitude longer than its typical trades. Furthermore, the model's prediction values reached extreme levels, far outside their normal stationary range, indicating that the magnitude of the order book imbalance was unprecedented. The model correctly identified the direction of the move but was, in a sense, overwhelmed by the conclusiveness of the signal.

While profitable, this event highlights a darker, reflexive nature of such algorithmic strategies. If a significant portion of market participants were to deploy similar imbalance-based models, it could create a dangerous feedback loop. An initial price dip would create an imbalance, triggering algorithmic selling, which in turn would exacerbate the imbalance, leading to more selling. This self-reinforcing cascade can transform a minor fluctuation into a full-blown flash crash. This mechanism is reminiscent of the models of predatory trading, such as that of Carlin, Sousa Lobo, and Viswanathan \citep{Carlin2007}, who demonstrate how strategic traders can exploit information in a manner that leads to episodic liquidity crises and contagion. In the context of our model, the "predators" are not necessarily fundamentally informed but are reacting to the mechanical signatures of other algorithms, creating a fragile market ecology.

\begin{figure}[h!]
    \centering
    \includegraphics[width=\textwidth]{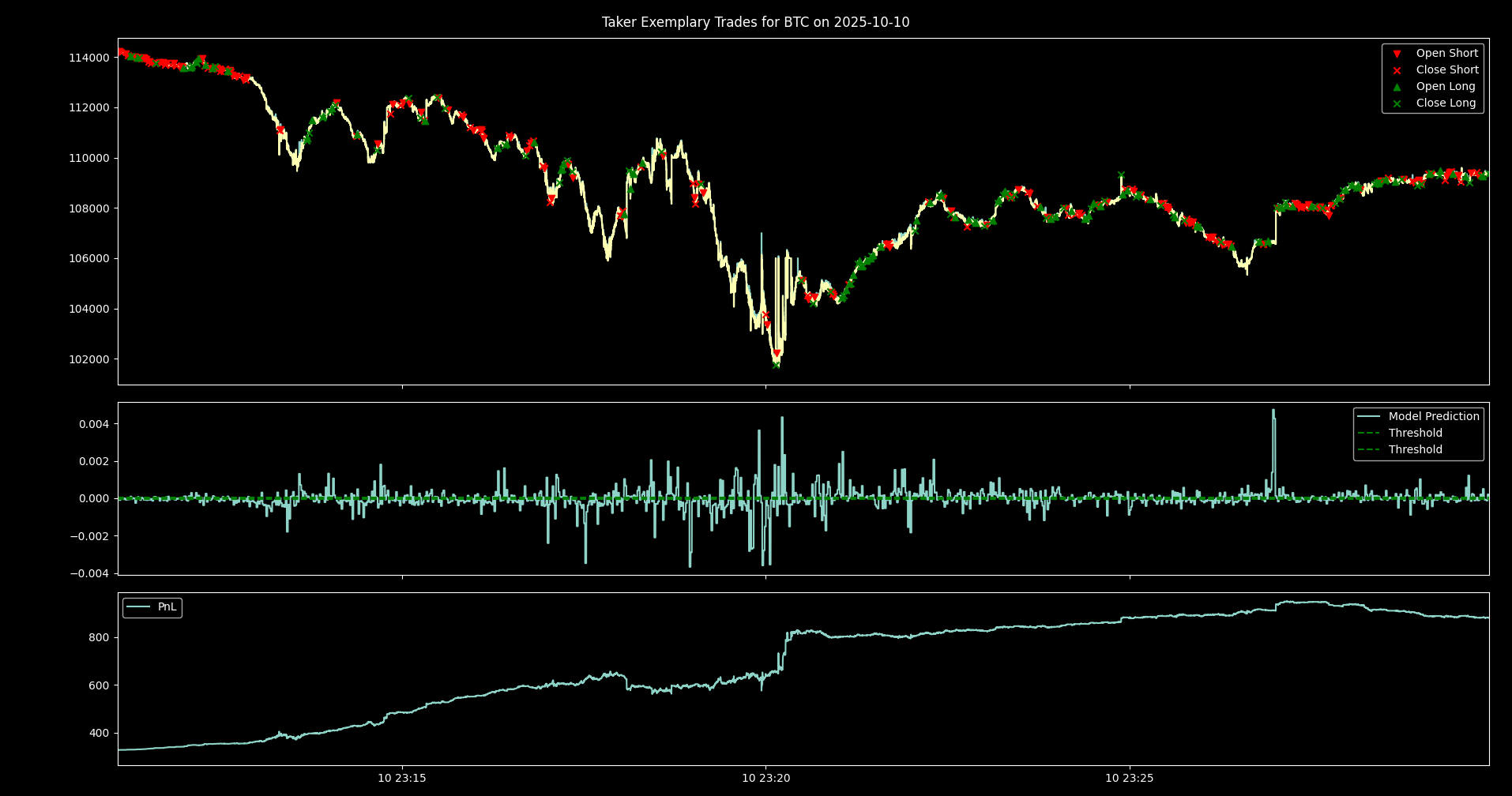}
    \caption{A detailed view of the flash crash, showing the model's reaction to a large, shallow bid order amidst the downturn.}
    \label{fig:taker_fc2}
\end{figure}

Figure \ref{fig:taker_fc2} provides a more granular view of the crash dynamics. At approximately 23:27, a large bid order was placed shallowly in the book, causing a momentary, sharp upward prediction from our model, which it immediately acted upon. This illustrates the model's high sensitivity to large, immediate changes in the order book. It also underscores the potential for market manipulation; a single, large, non-bona fide order could theoretically trigger a cascade of algorithmic trades.

\subsection{Implications for Market Stability and Policy}

The success of our taker strategy during the flash crash, while a testament to its predictive power, raises important questions about the social value of such automated trading and its impact on market organization. A market dominated by high-frequency strategies predicated on the same signals (like order book imbalance) is inherently fragile. It fosters a technological "arms race" where speed is the only determinant of profitability, and the first few participants to react capture all the gains, while the ensuing cascade degrades market quality for everyone else.

From a policy perspective, these findings suggest that regulators should consider the systemic risks posed by the homogenization of algorithmic strategies. Potential interventions could include dynamic "circuit breakers" triggered by extreme order imbalance, modifications to the market's tick size, or even imposing a minimum resting time for orders to discourage fleeting, predatory liquidity. Ultimately, while our model demonstrates that it is possible to build a profitable trading strategy based on market microstructure signals, its widespread adoption could inadvertently undermine the very stability and liquidity it seeks to exploit, leading to a market that is less resilient and more susceptible to systemic shocks.

\subsection{Market-Making Under Extreme Conditions: A Cautionary Tale}

\begin{figure}[h!]
    \centering
    \includegraphics[width=\textwidth]{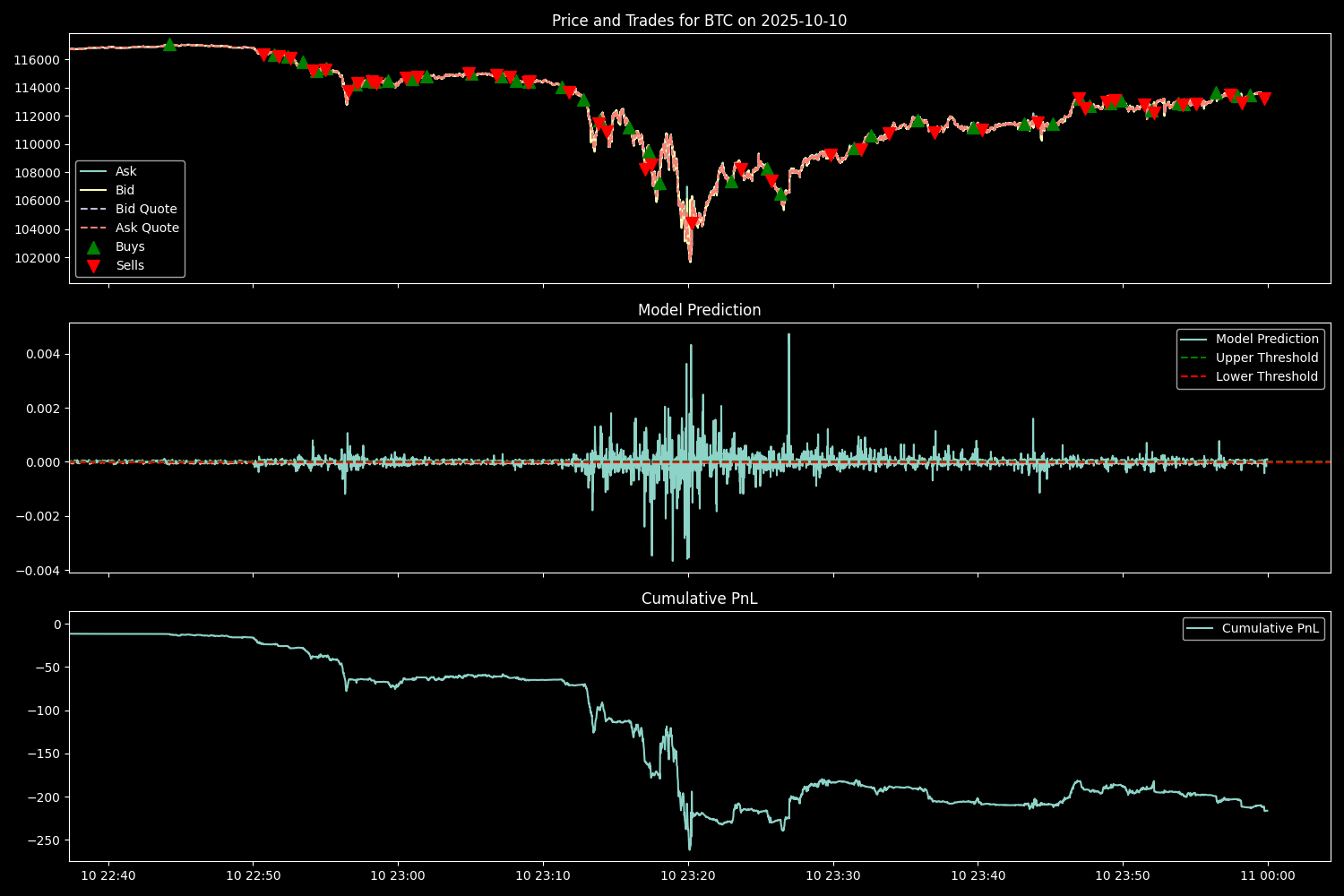}
    \caption{Market-maker strategy performance during the October 10, 2025 flash crash. The strategy repeatedly gets filled on the bid side, accumulating a losing long position and suffering significant losses.}
    \label{fig:maker_fc}
\end{figure}

In stark contrast to the taker strategy's success, the market-making strategy suffered catastrophic losses during the flash crash, as illustrated in Figure \ref{fig:maker_fc}. The strategy, which operates by posting passive bid and ask quotes around the mid-price, is designed to profit from the bid-ask spread in normal market conditions. However, during the unidirectional price collapse on October 10, it fell victim to severe adverse selection.

As the market plummeted, the strategy was repeatedly filled on its bid-side quotes, forcing it to accumulate a growing, and increasingly unprofitable, long position. Because the price was falling so rapidly, there was no corresponding fill on the ask side to offset the risk. The PnL curve shows a steep decline, culminating in a crash around 23:20, wiping out thousands of dollars in simulated equity. The model's predictions, as seen in the middle panel, were orders of magnitude larger than normal, yet the maker strategy's simple, constant-spread logic was unable to adapt to this extreme signal.

This failure is a textbook illustration of classic market microstructure theories, such as the models of Glosten and Milgrom (1985) and Kyle (1985). These models posit that the bid-ask spread is not just a source of profit for liquidity providers, but a crucial form of compensation for the risk of trading against informed participants---a phenomenon known as adverse selection. During periods of high volatility and strong directional conviction, like a flash crash, the probability of trading against informed flow increases dramatically. A market maker who fails to widen their spread in response is essentially offering a subsidy to informed traders, leading to near-certain losses. Our maker strategy's failure stems directly from its inability to dynamically adjust its spread to account for this surge in adverse selection risk.

\begin{figure}[h!]
    \centering
    \includegraphics[width=\textwidth]{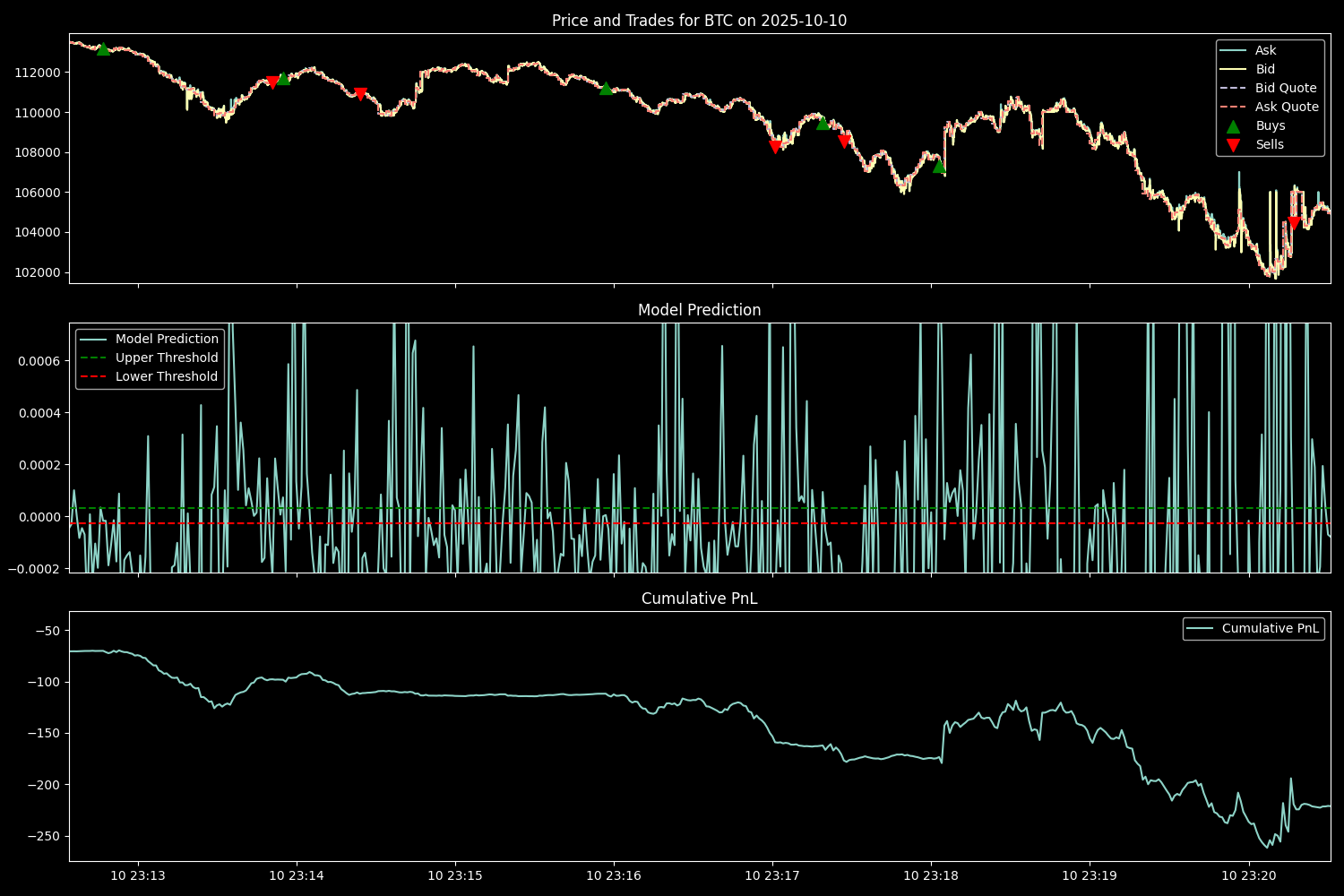}
    \caption{A detailed view of the market-maker's activity, showing the persistent fills on the bid side against a backdrop of extreme model predictions.}
    \label{fig:maker_fc2}
\end{figure}

The juxtaposition of the two strategies is telling. The taker strategy profited because it acted as an informed agent, using the order book imbalance to correctly predict the market's direction and take liquidity. The maker strategy lost money precisely because it was providing liquidity to this informed, directional flow. This reveals a fundamental tension in market microstructure: during periods of turmoil, liquidity-taking (if directionally correct) is profitable, while naive liquidity-providing is perilous. The taker strategy won by exploiting the very information that made the maker strategy insolvent. This underscores that a successful market-making algorithm must not only predict short-term price movements but also dynamically manage its spreads to survive periods of high information asymmetry.
\section{Conclusion}
The empirical invariances documented here point toward a scale-free description of short-horizon price formation. Expressed in relative prices and flows, the mapping from microstructure state to returns appears stable across market capitalization tiers. This aligns with theories in which liquidity provision internalizes inventory risk and adverse selection similarly irrespective of asset size, once scaled by local liquidity and tick structure.

Interpretability gains from SHAP are substantive. Importance hierarchies validate that a small set of features,order flow imbalance, spreads, and VWAP deviations,drive most predictive power, while dependence shapes quantify monotonicities, asymmetries, and saturation. These diagnostics help discriminate genuine economic channels from dataset-specific proxies and provide actionable guidance for risk controls (e.g., deactivating signals in wide-spread regimes) and execution design.

Our primary novelty and contribution to existing literature is the robustness analysis of the model during the major flash crash, where we validate whether the patterns observed in normal market conditions persist under tail events. We find that the taker strategy performs well during the flash crash, while the maker strategy suffers significant losses. This suggests that the taker strategy is more robust to extreme market conditions, while the maker strategy is more vulnerable.

From a deployment perspective, universal feature libraries enable rapid porting of predictive models across assets with minimal re-engineering. The observed concavity in dependence suggests diminishing marginal returns to liquidity-taking pressure, arguing for conservative sizing at extremes and caution when extrapolating beyond the observed domain. The robustness of results across objectives and horizons further supports operational reliability.

Limitations include the focus on a single short horizon, potential venue- or regime-specific effects, and the use of a single model family. Extending to cross-venue consolidation, dynamic horizons, and alternative interpretable models would strengthen external validity. Causal identification of order flow effects remains an open avenue, potentially via instrumental strategies or market design changes that induce exogenous variation in liquidity and flow.

We demonstrate that a compact set of order book and trade features yields similar predictive importance and SHAP dependence shapes across cryptocurrencies spanning market capitalization tiers. This universality aligns with microstructure theory and supports the portability of feature libraries across assets.

Under a conservative taker execution, forecasts translate into tradable signals. Robustness checks confirm stability to thresholds, fees, splits, and objectives. We view these results as evidence for shared microstructure features and a foundation for scalable, cross-asset short-horizon modeling in crypto markets.

\newpage
\section*{Appendix A: Supplementary Charts}

\begin{figure}[H]
    \centering
    \begin{subfigure}{0.19\textwidth}
      \includegraphics[width=\linewidth]{charts/BTC/feature1.png}
      \caption{BTC - L1 Imbalance}
    \end{subfigure}
    \begin{subfigure}{0.19\textwidth}
      \includegraphics[width=\linewidth]{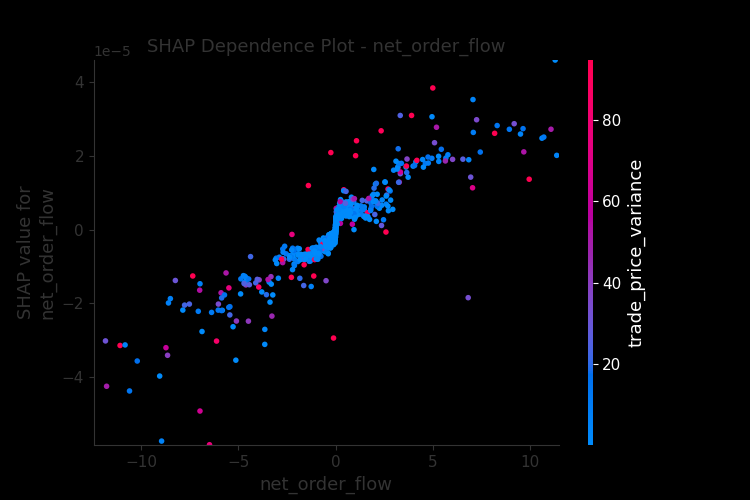}
      \caption{BTC - Net Order Flow}
    \end{subfigure}
    \begin{subfigure}{0.19\textwidth}
      \includegraphics[width=\linewidth]{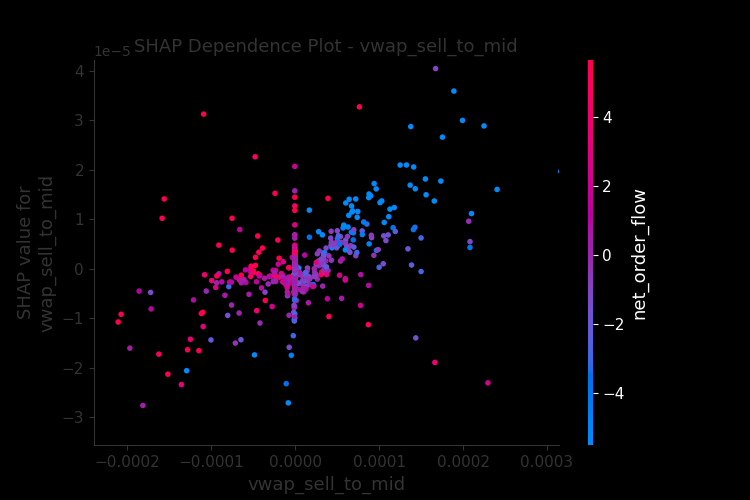}
      \caption{BTC - Vwap Sell to Mid}
    \end{subfigure}
    \begin{subfigure}{0.19\textwidth}
      \includegraphics[width=\linewidth]{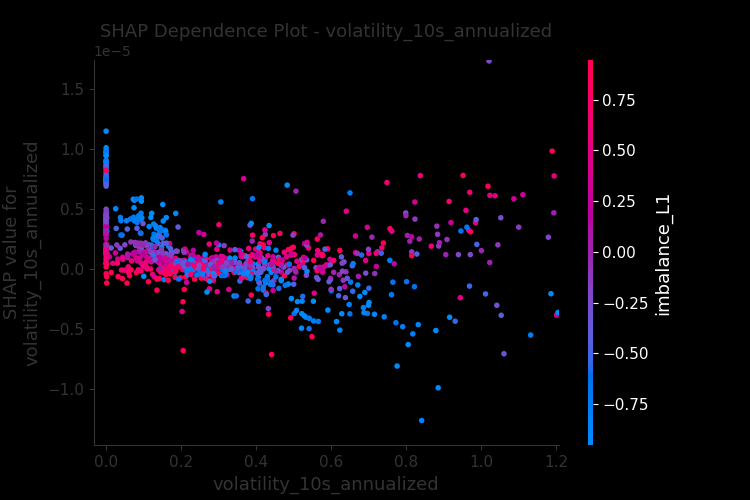}
      \caption{BTC - Volatility}
    \end{subfigure}
    \begin{subfigure}{0.19\textwidth}
      \includegraphics[width=\linewidth]{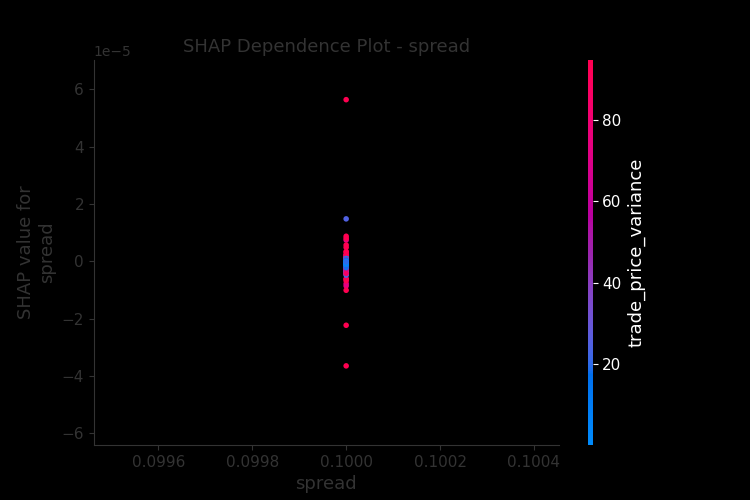}
      \caption{BTC - Spread}
    \end{subfigure}
    \begin{subfigure}{0.19\textwidth}
      \includegraphics[width=\linewidth]{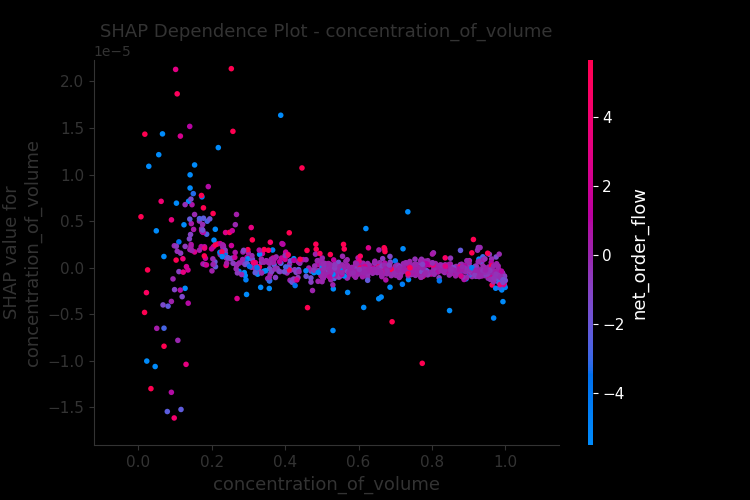}
      \caption{BTC - Volume Concentration}
    \end{subfigure}
    \begin{subfigure}{0.19\textwidth}
      \includegraphics[width=\linewidth]{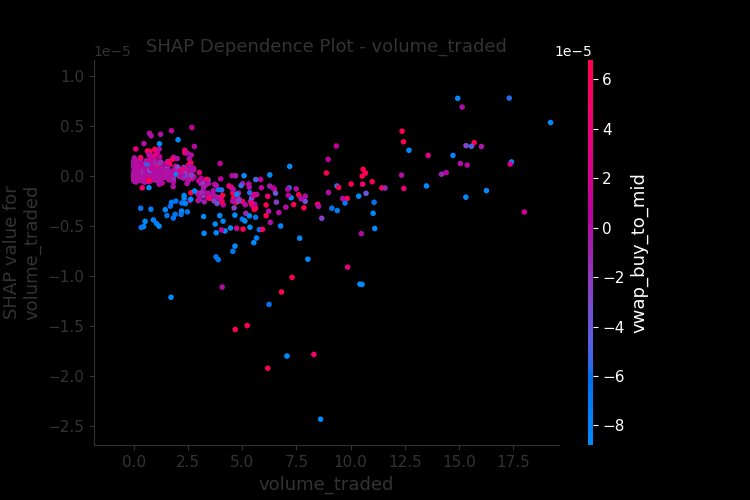}
      \caption{BTC - Volume Traded}
    \end{subfigure}
    \begin{subfigure}{0.19\textwidth}
      \includegraphics[width=\linewidth]{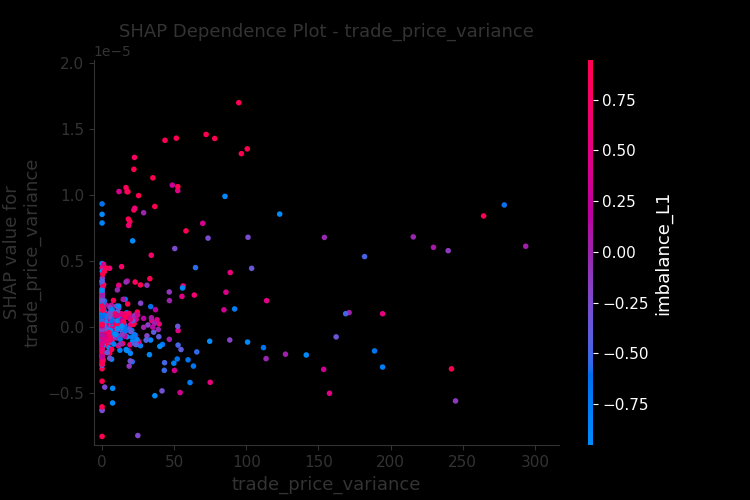}
      \caption{BTC - Trade Price Variance}
    \end{subfigure}
    \begin{subfigure}{0.19\textwidth}
      \includegraphics[width=\linewidth]{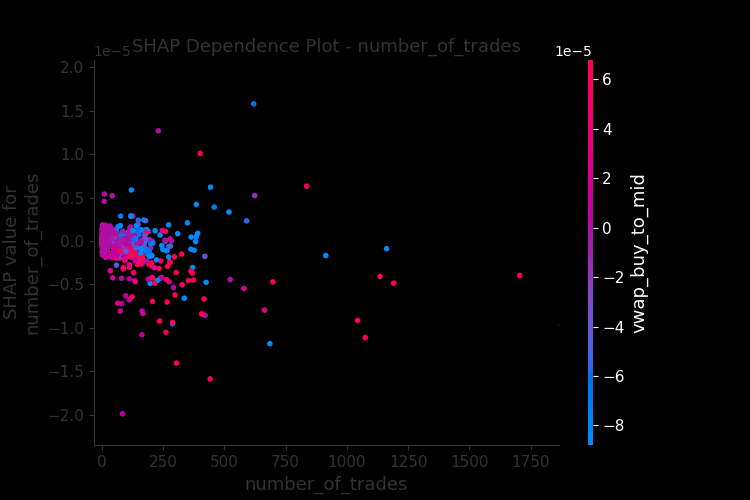}
      \caption{BTC - N Trades}
    \end{subfigure}
    \begin{subfigure}{0.19\textwidth}
      \includegraphics[width=\linewidth]{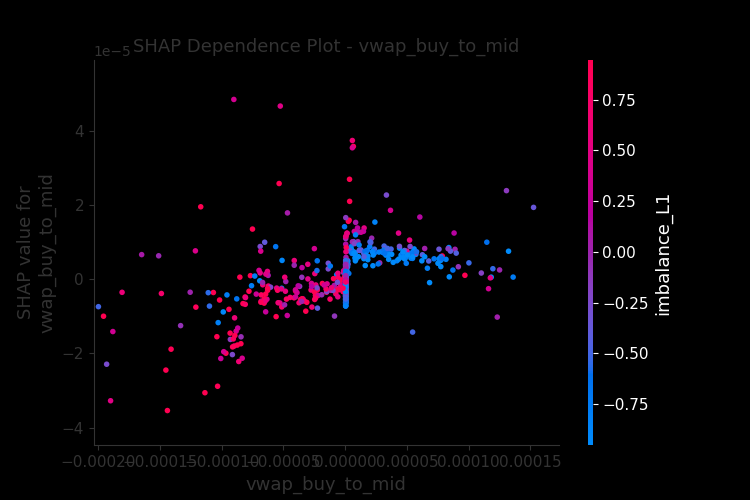}
      \caption{BTC - VWAP Buy to Mid}
    \end{subfigure}
    \caption{BTC SHAP dependence plots (features 1–10).}
  \end{figure}
  
  \begin{figure}[H]
    \centering
    \begin{subfigure}{0.19\textwidth}
      \includegraphics[width=\linewidth]{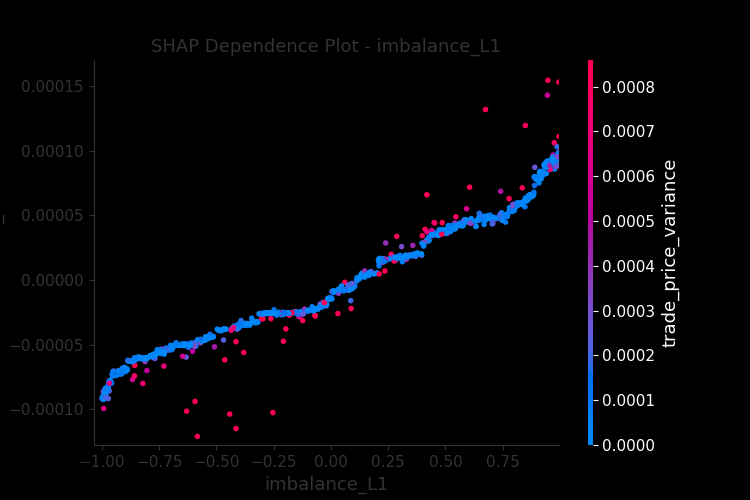}
      \caption{LTC - L1 Imbalance}
    \end{subfigure}
    \begin{subfigure}{0.19\textwidth}
      \includegraphics[width=\linewidth]{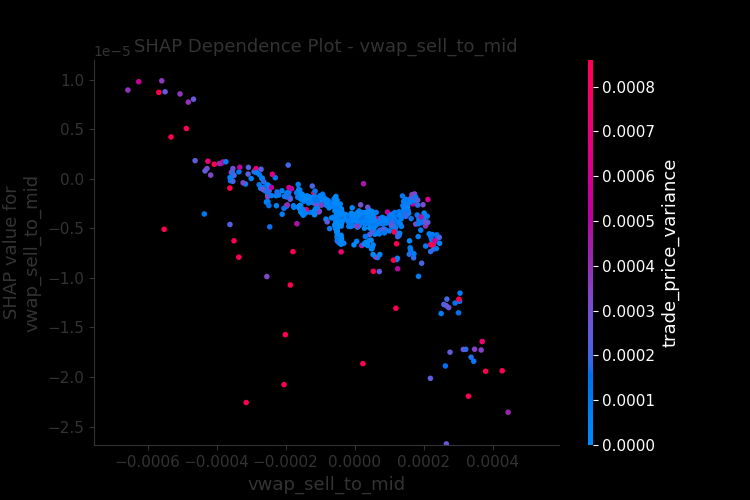}
      \caption{LTC - VWAP Sell to Mid}
    \end{subfigure}
    \begin{subfigure}{0.19\textwidth}
      \includegraphics[width=\linewidth]{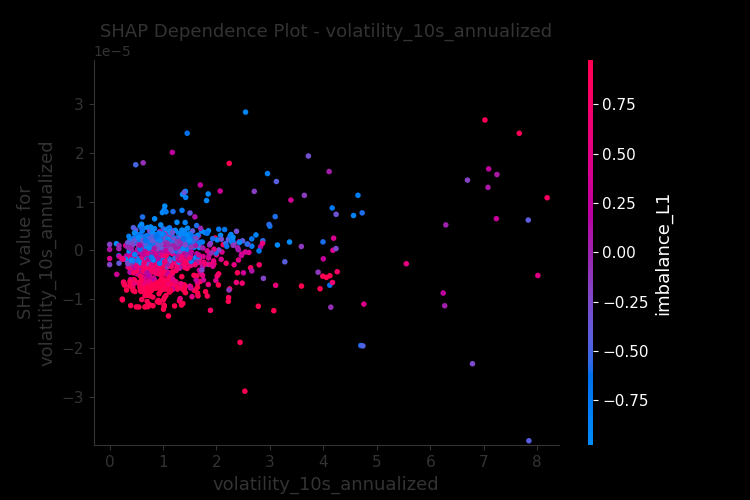}
      \caption{LTC - Volatility}
    \end{subfigure}
    \begin{subfigure}{0.19\textwidth}
      \includegraphics[width=\linewidth]{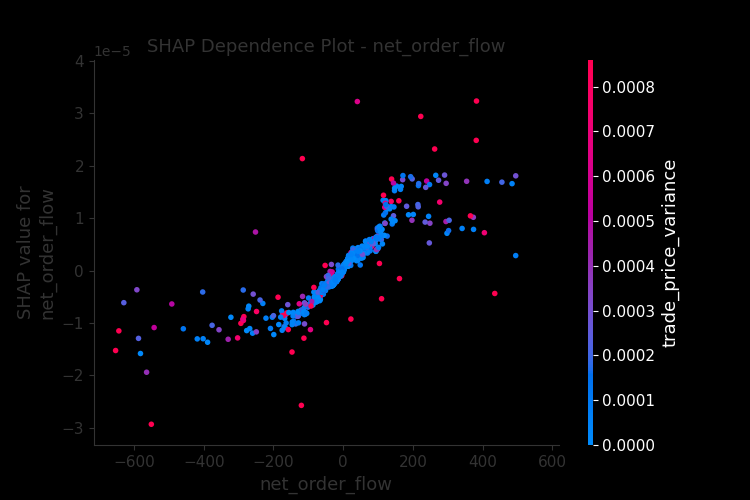}
      \caption{LTC - Net Order Flow}
    \end{subfigure}
    \begin{subfigure}{0.19\textwidth}
      \includegraphics[width=\linewidth]{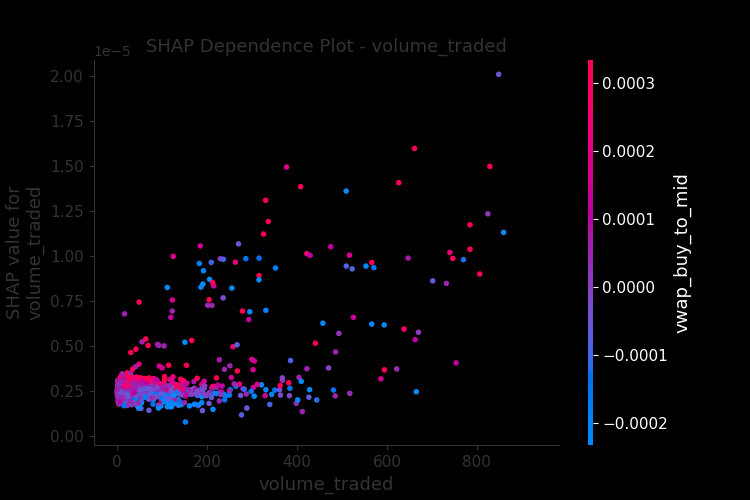}
      \caption{LTC - Volume Traded}
    \end{subfigure}
    \begin{subfigure}{0.19\textwidth}
      \includegraphics[width=\linewidth]{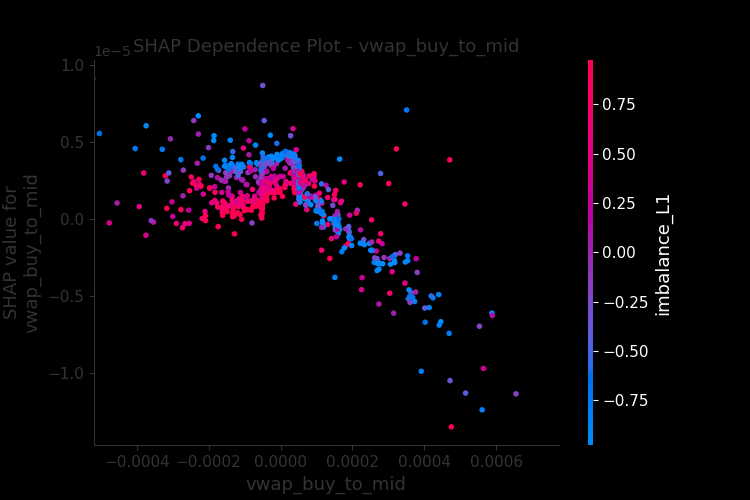}
      \caption{LTC - VWAP Buy to Mid}
    \end{subfigure}
    \begin{subfigure}{0.19\textwidth}
      \includegraphics[width=\linewidth]{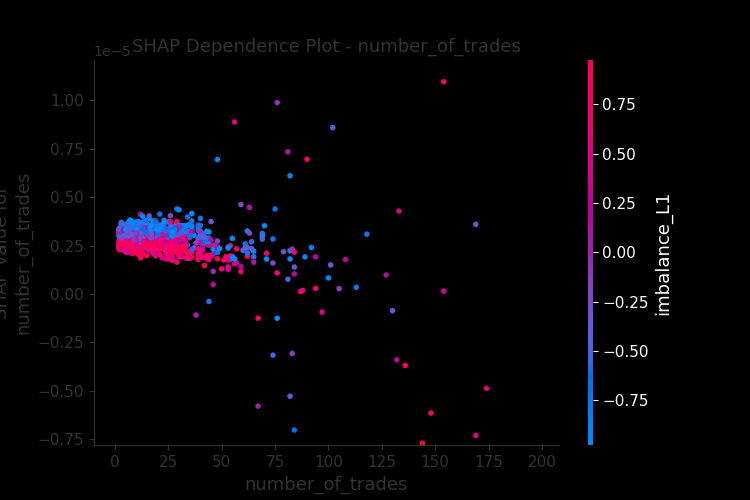}
      \caption{LTC - N Trades}
    \end{subfigure}
    \begin{subfigure}{0.19\textwidth}
      \includegraphics[width=\linewidth]{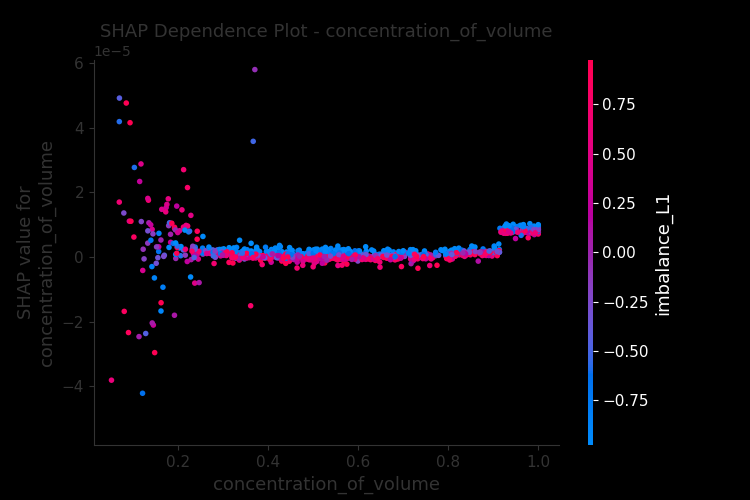}
      \caption{LTC - Concentration of Volume}
    \end{subfigure}
    \begin{subfigure}{0.19\textwidth}
      \includegraphics[width=\linewidth]{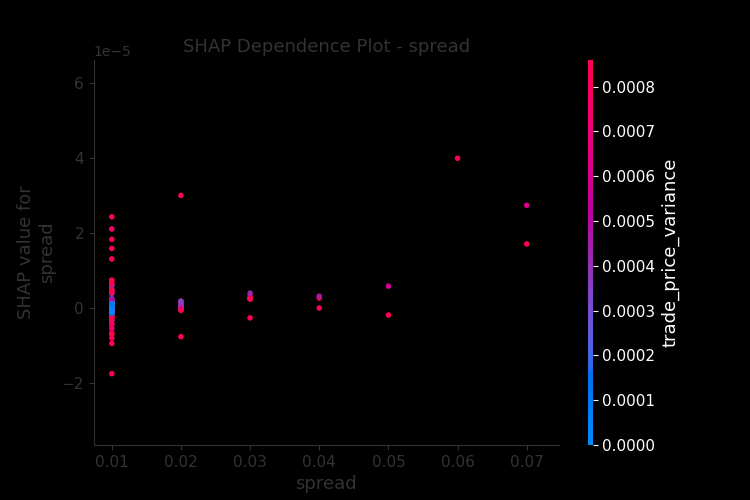}
      \caption{LTC - Spread}
    \end{subfigure}
    \begin{subfigure}{0.19\textwidth}
      \includegraphics[width=\linewidth]{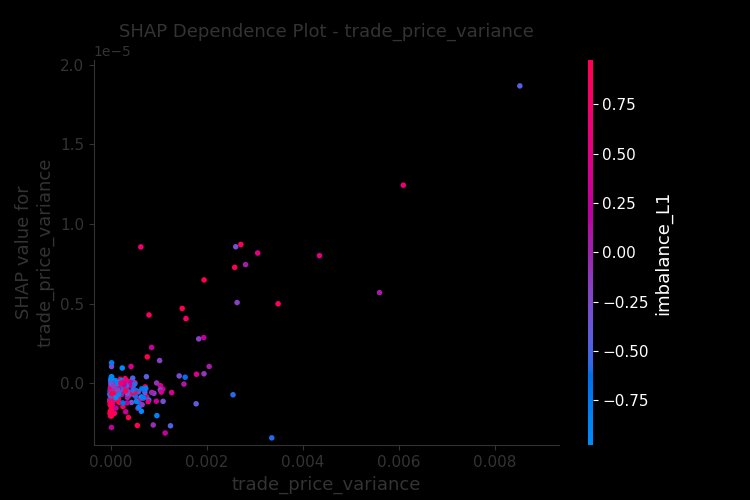}
      \caption{LTC - Trade Price Variance}
    \end{subfigure}
    \caption{LTC SHAP dependence plots (features 1–10).}
  \end{figure}
  
  \begin{figure}[H]
    \centering
    \begin{subfigure}{0.19\textwidth}
      \includegraphics[width=\linewidth]{charts/ETC/feature1.png}
      \caption{ETC - L1 Imbalance}
    \end{subfigure}
    \begin{subfigure}{0.19\textwidth}
      \includegraphics[width=\linewidth]{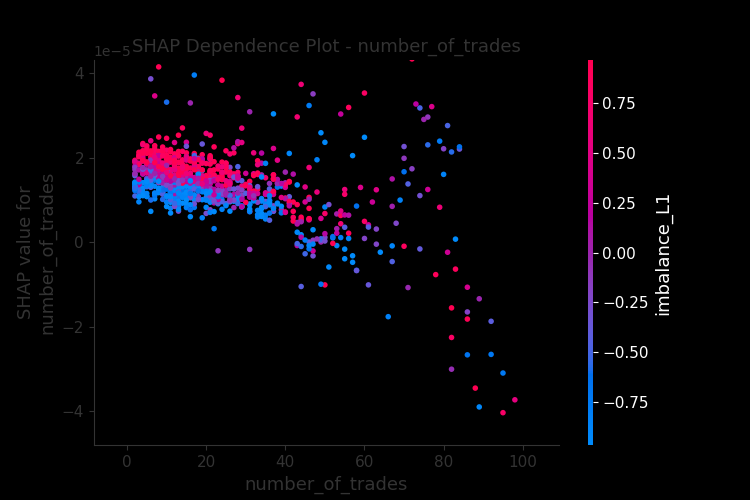}
      \caption{ETC - N Trades}
    \end{subfigure}
    \begin{subfigure}{0.19\textwidth}
      \includegraphics[width=\linewidth]{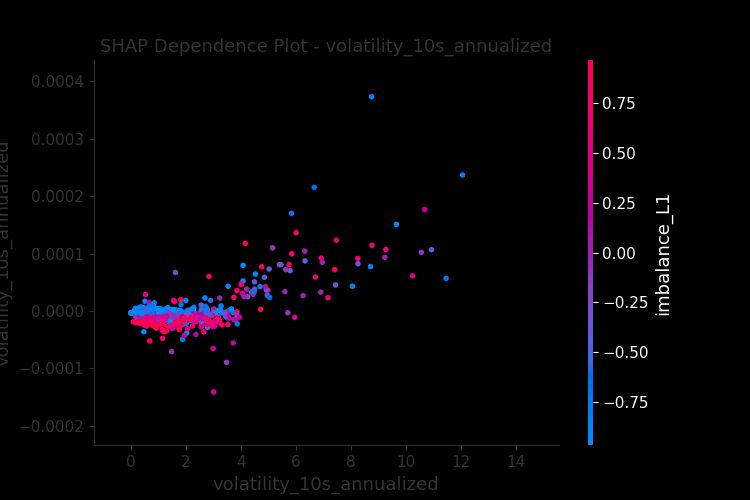}
      \caption{ETC - Volatility}
    \end{subfigure}
    \begin{subfigure}{0.19\textwidth}
      \includegraphics[width=\linewidth]{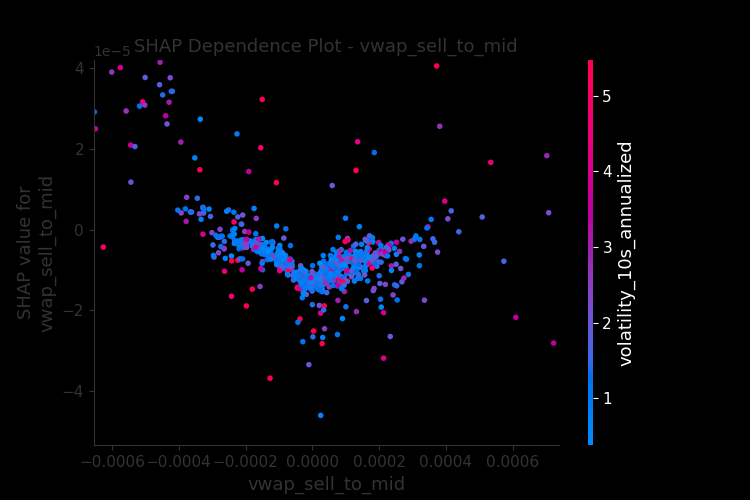}
      \caption{ETC - VWAP Sell to Mid}
    \end{subfigure}
    \begin{subfigure}{0.19\textwidth}
      \includegraphics[width=\linewidth]{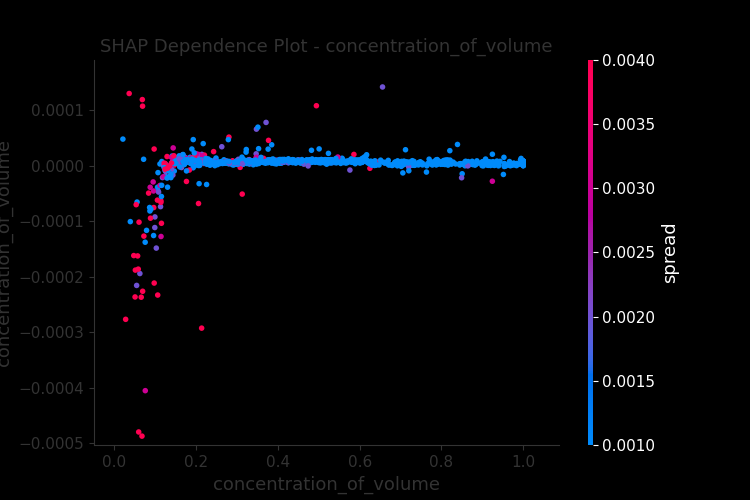}
      \caption{ETC - Concentration of Volume}
    \end{subfigure}
    \begin{subfigure}{0.19\textwidth}
      \includegraphics[width=\linewidth]{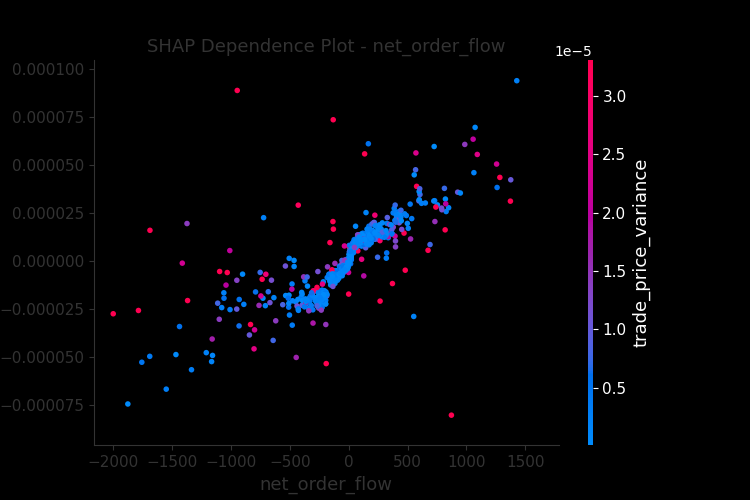}
      \caption{ETC - Net Order Flow}
    \end{subfigure}
    \begin{subfigure}{0.19\textwidth}
      \includegraphics[width=\linewidth]{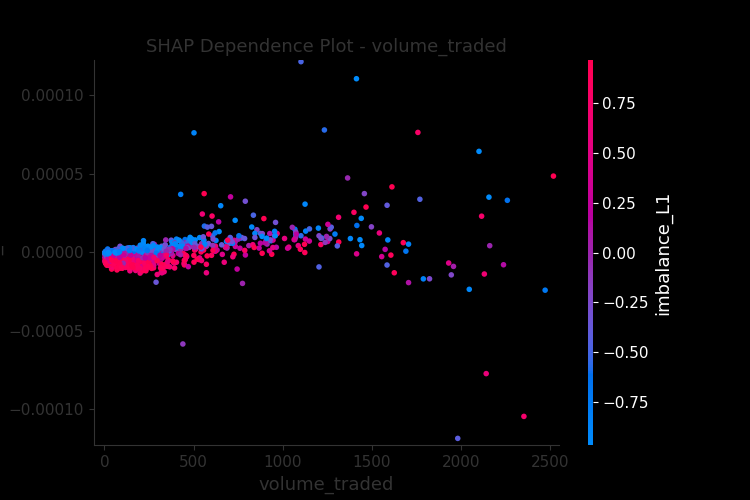}
      \caption{ETC - Volume Traded}
    \end{subfigure}
    \begin{subfigure}{0.19\textwidth}
      \includegraphics[width=\linewidth]{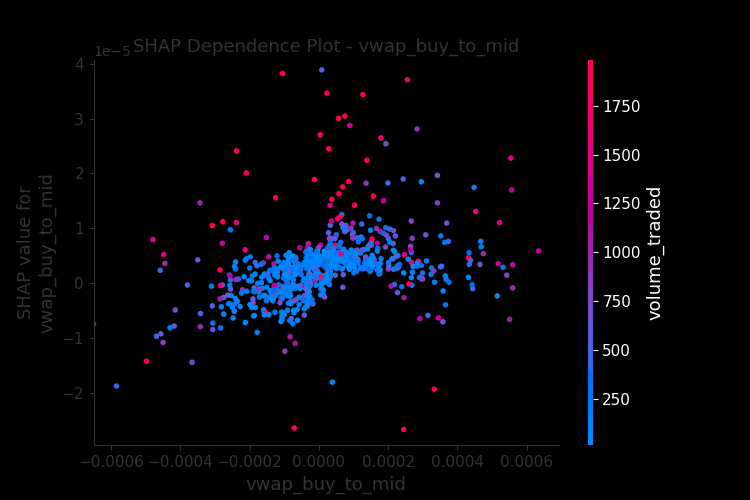}
      \caption{ETC - VWAP Buy to Mid}
    \end{subfigure}
    \begin{subfigure}{0.19\textwidth}
      \includegraphics[width=\linewidth]{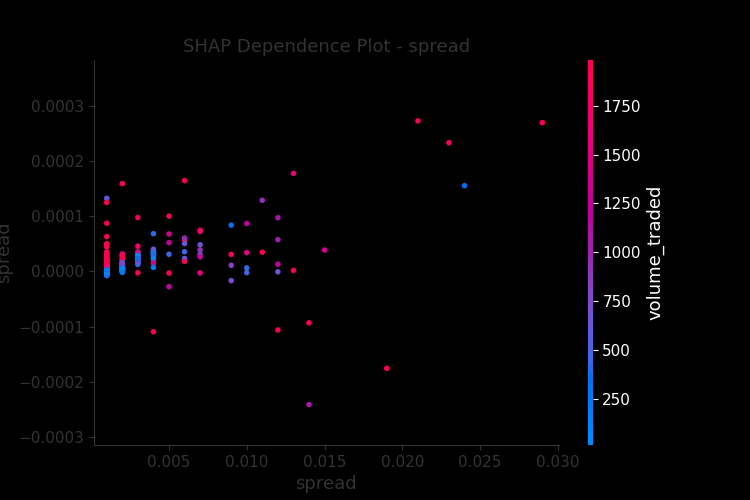}
      \caption{ETC - Spread}
    \end{subfigure}
    \begin{subfigure}{0.19\textwidth}
      \includegraphics[width=\linewidth]{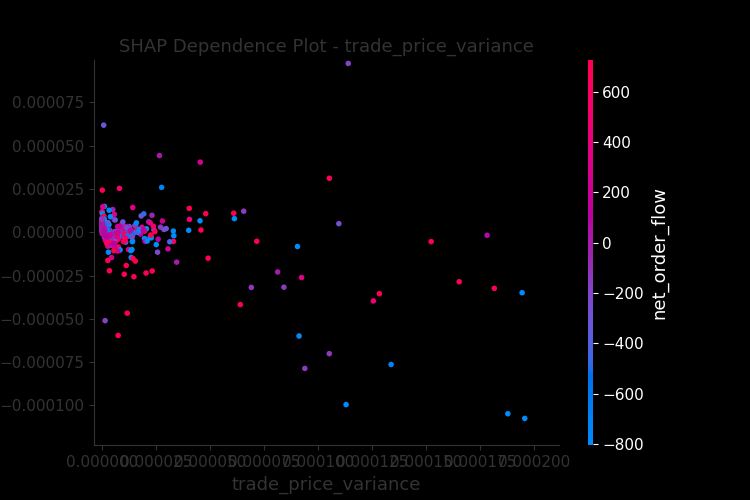}
      \caption{ETC - Trade Price Variance}
    \end{subfigure}
    \caption{ETC SHAP dependence plots (features 1–10).}
  \end{figure}
  
  \begin{figure}[H]
    \centering
    \begin{subfigure}{0.19\textwidth}
      \includegraphics[width=\linewidth]{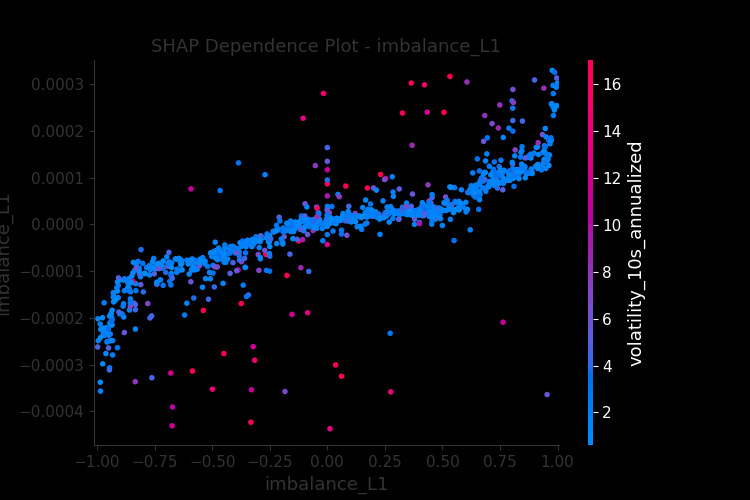}
      \caption{ENJ - L1 Imbalance}
    \end{subfigure}
    \begin{subfigure}{0.19\textwidth}
      \includegraphics[width=\linewidth]{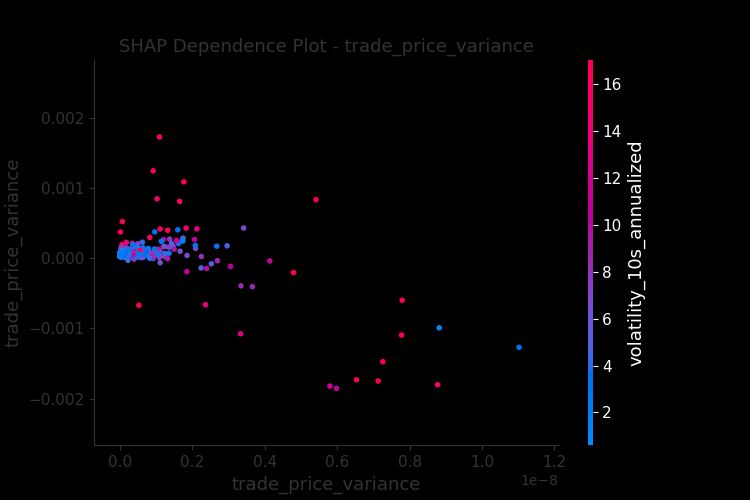}
      \caption{ENJ - Trade Price Variance}
    \end{subfigure}
    \begin{subfigure}{0.19\textwidth}
      \includegraphics[width=\linewidth]{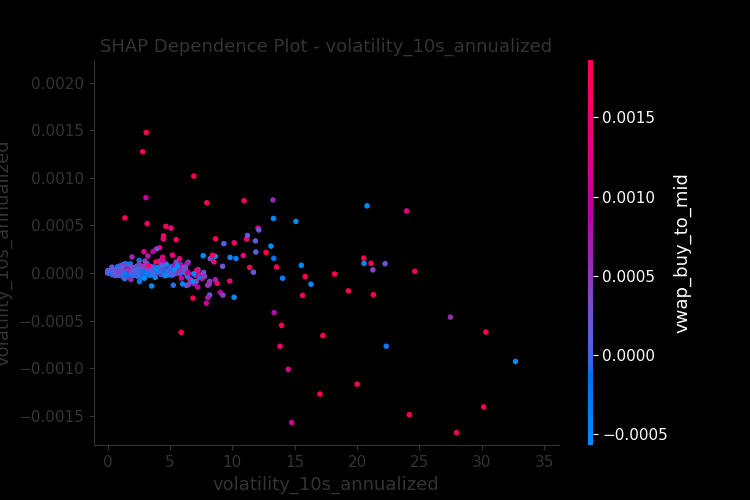}
      \caption{ENJ - Volatility}
    \end{subfigure}
    \begin{subfigure}{0.19\textwidth}
      \includegraphics[width=\linewidth]{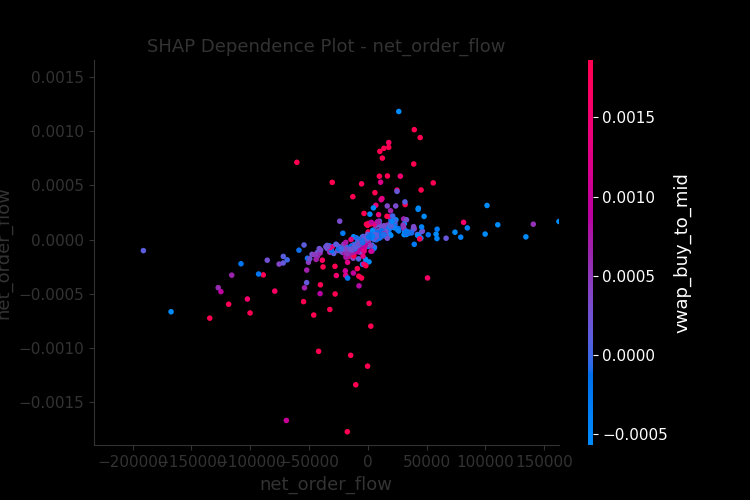}
      \caption{ENJ - Net Order Flow}
    \end{subfigure}
    \begin{subfigure}{0.19\textwidth}
      \includegraphics[width=\linewidth]{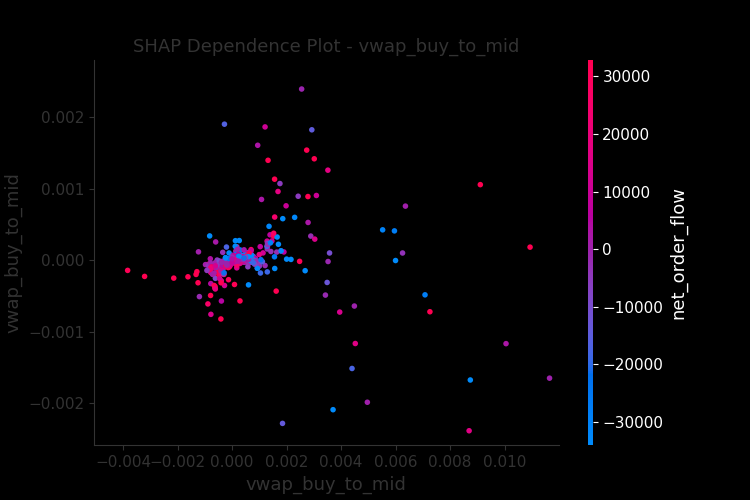}
      \caption{ENJ - VWAP Buy to Mid}
    \end{subfigure}
    \begin{subfigure}{0.19\textwidth}
      \includegraphics[width=\linewidth]{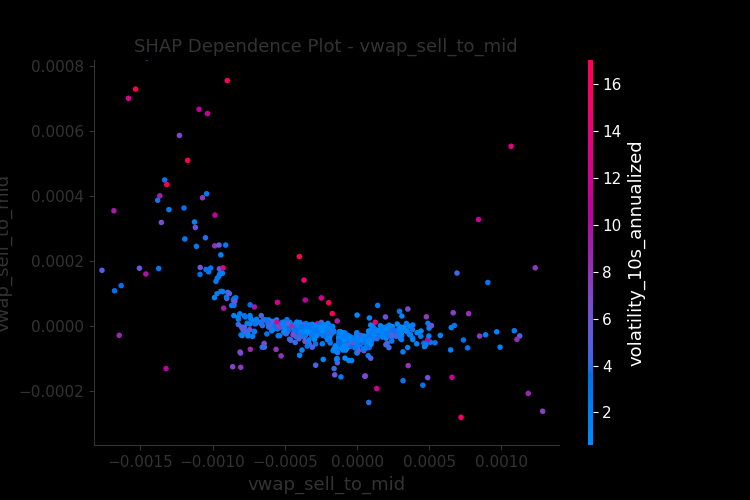}
      \caption{ENJ - VWAP Sell to Mid}
    \end{subfigure}
    \begin{subfigure}{0.19\textwidth}
      \includegraphics[width=\linewidth]{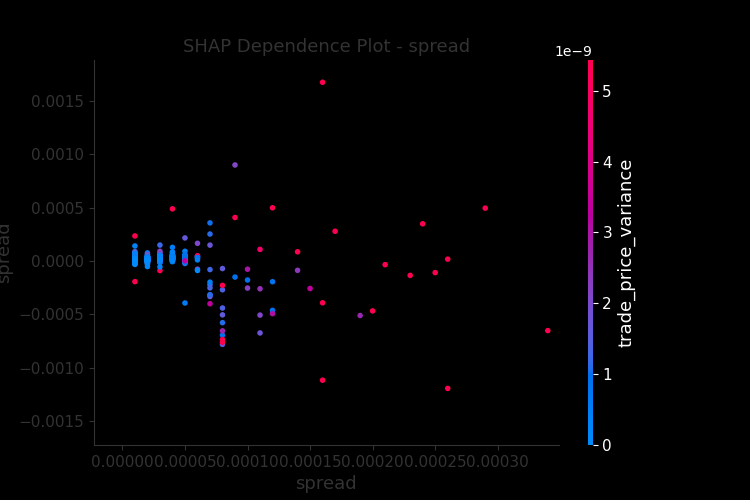}
      \caption{ENJ - Spread}
    \end{subfigure}
    \begin{subfigure}{0.19\textwidth}
      \includegraphics[width=\linewidth]{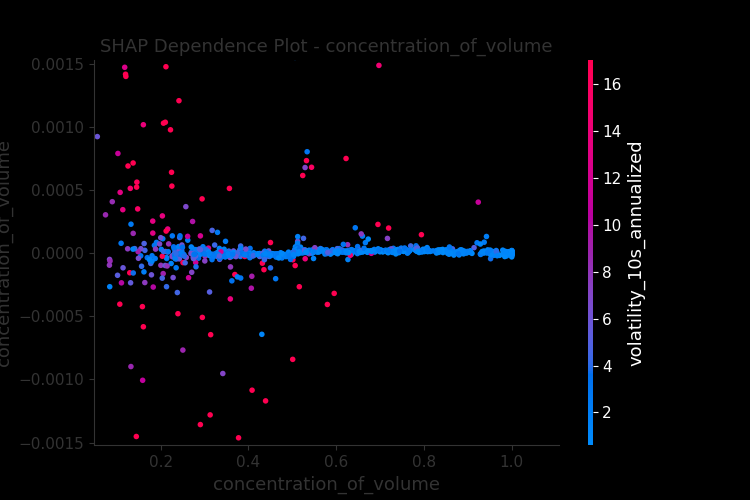}
      \caption{ENJ - Concentration of Volume}
    \end{subfigure}
    \begin{subfigure}{0.19\textwidth}
      \includegraphics[width=\linewidth]{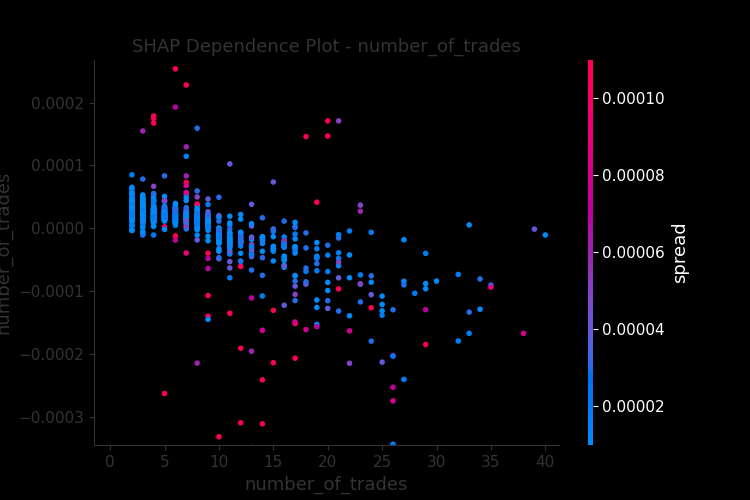}
      \caption{ENJ - N Trades}
    \end{subfigure}
    \begin{subfigure}{0.19\textwidth}
      \includegraphics[width=\linewidth]{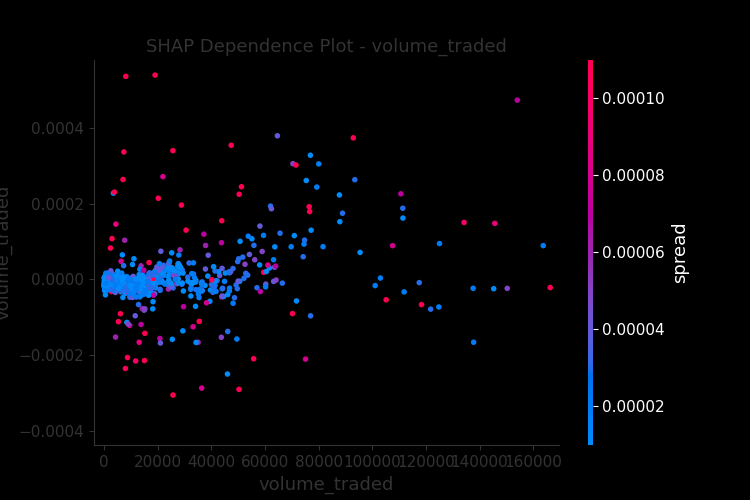}
      \caption{ENJ - Volume Traded}
    \end{subfigure}
    \caption{ENJ SHAP dependence plots (features 1–10).}
  \end{figure}
  
  \begin{figure}[H]
    \centering
    \begin{subfigure}{0.19\textwidth}
      \includegraphics[width=\linewidth]{charts/ROSE/feature1.png}
      \caption{ROSE - Imbalance}
    \end{subfigure}
    \begin{subfigure}{0.19\textwidth}
      \includegraphics[width=\linewidth]{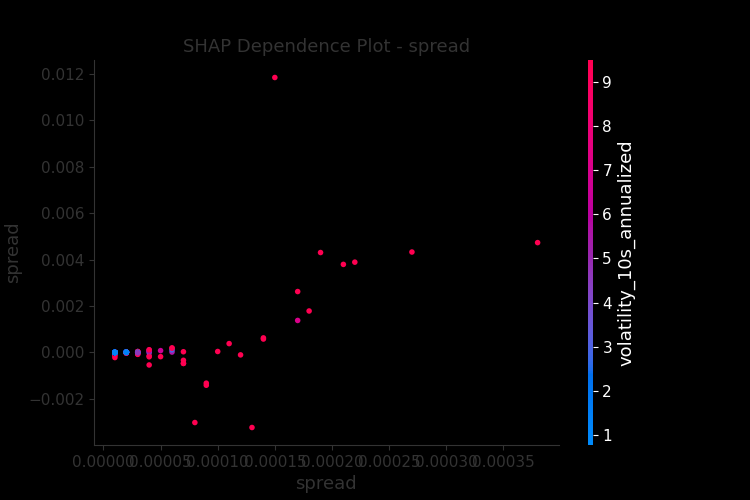}
      \caption{ROSE - Spread}
    \end{subfigure}
    \begin{subfigure}{0.19\textwidth}
      \includegraphics[width=\linewidth]{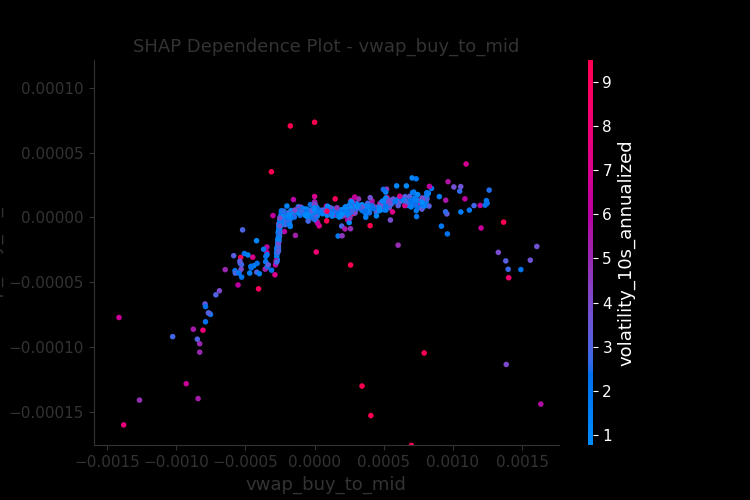}
      \caption{ROSE - Vwap Buy to Mid}
    \end{subfigure}
    \begin{subfigure}{0.19\textwidth}
      \includegraphics[width=\linewidth]{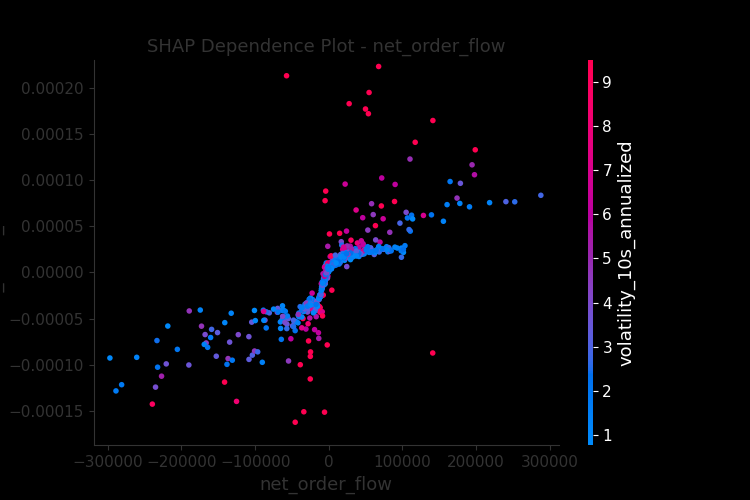}
      \caption{ROSE - Net Order Flow}
    \end{subfigure}
    \begin{subfigure}{0.19\textwidth}
      \includegraphics[width=\linewidth]{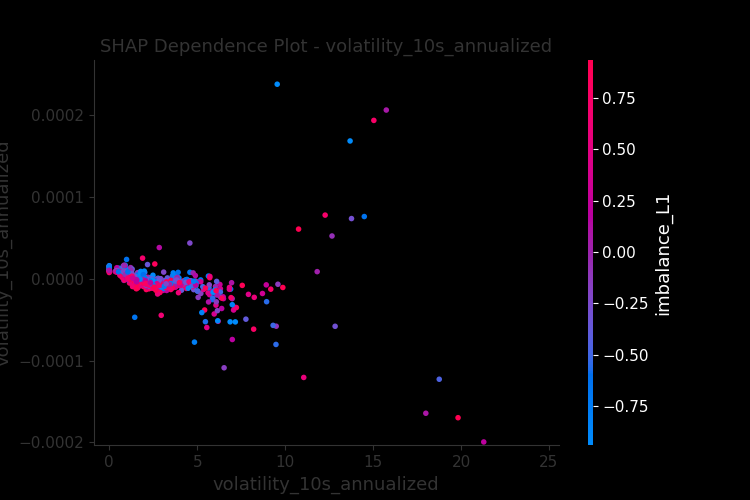}
      \caption{ROSE - Volatility}
    \end{subfigure}
    \begin{subfigure}{0.19\textwidth}
      \includegraphics[width=\linewidth]{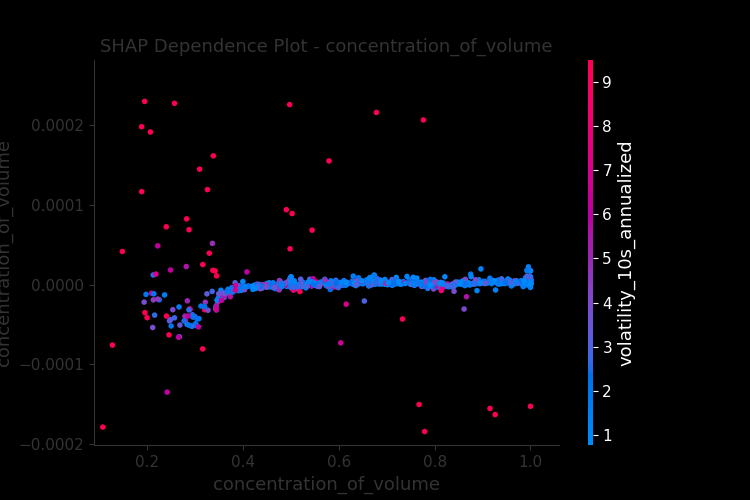}
      \caption{ROSE - Concentration of Volume}
    \end{subfigure}
    \begin{subfigure}{0.19\textwidth}
      \includegraphics[width=\linewidth]{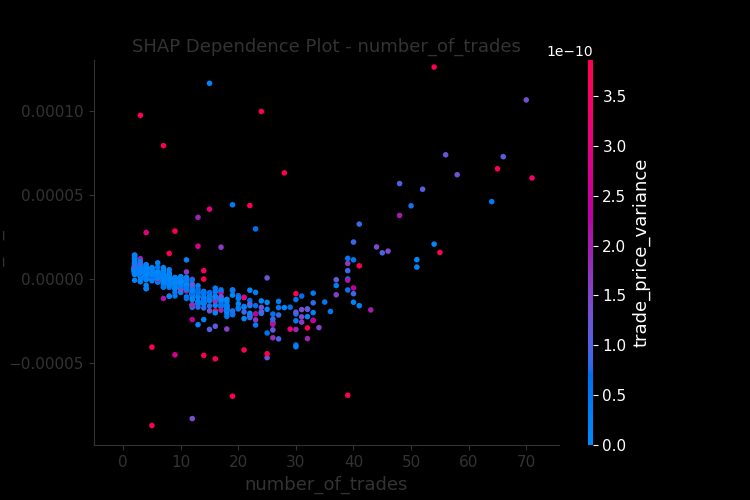}
      \caption{ROSE - N Trades}
    \end{subfigure}
    \begin{subfigure}{0.19\textwidth}
      \includegraphics[width=\linewidth]{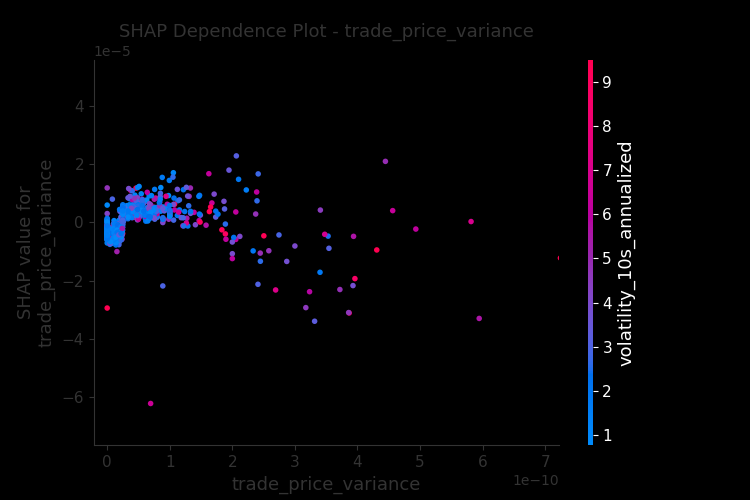}
      \caption{ROSE - Trade Price Variance}
    \end{subfigure}
    \begin{subfigure}{0.19\textwidth}
      \includegraphics[width=\linewidth]{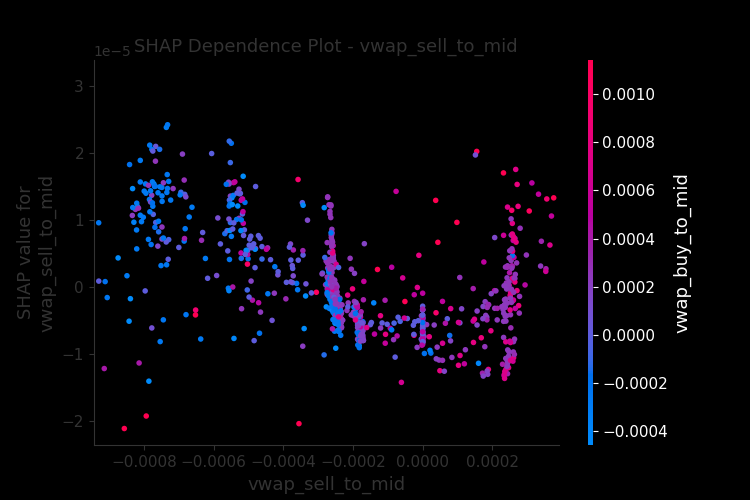}
      \caption{ROSE - VWAP Sell to Mid}
    \end{subfigure}
    \begin{subfigure}{0.19\textwidth}
      \includegraphics[width=\linewidth]{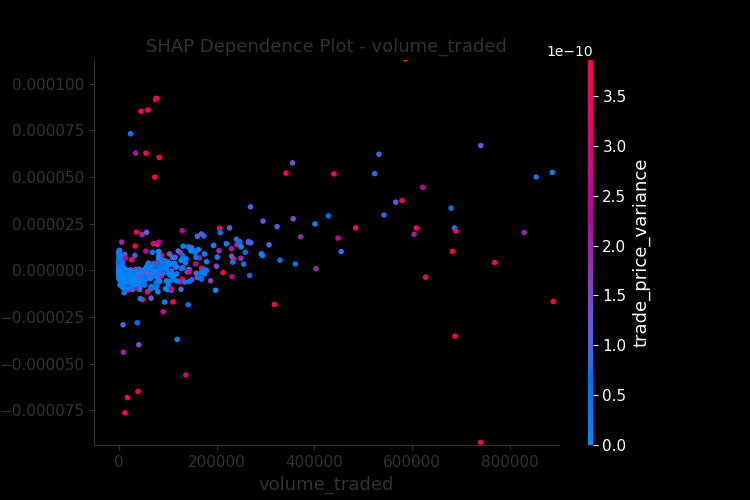}
      \caption{ROSE - Volume Traded}
    \end{subfigure}
    \caption{ROSE SHAP dependence plots (features 1–10).}
  \end{figure}
\newpage
\bibliographystyle{plainnat}
\bibliography{references}
\end{document}